\documentclass[aps,prd,12pt,notitlepage,nofootinbib,tightenlines]{revtex4-1}
\usepackage{amsmath}
\usepackage{bm}
\usepackage{times}
\usepackage{braket}
\usepackage{color}
\usepackage{epsfig}
\usepackage{slashed}
\usepackage{hyperref}
\usepackage{subfigure}
\usepackage{rotating}
\allowdisplaybreaks[1]
\newcommand{\beq}{\begin{eqnarray}}
\newcommand{\eeq}{\end{eqnarray}}
\newcommand{\non}{\nonumber\\ }

\newcommand{\bbc}{ {\cal B}_{bc}}
\newcommand{\bb}{ {\cal B}_b }

\newcommand{\psl}{ p \hspace{-1.8truemm}/ }
\newcommand{\ksl}{ k \hspace{-2truemm}/ }

\definecolor{Red}{rgb}{1.,0.,0.}

\definecolor{Blue}{rgb}{0.,0.,1.}

\definecolor{nicered}{rgb}{0.7,0.1,0.1}
\definecolor{nicegreen}{rgb}{0.1,0.5,0.1}

\hypersetup{colorlinks,citecolor=nicegreen,linkcolor=nicered}

\begin{document}
\title{ Weak decays of bottom-charm baryons: $\mathcal{B}_{bc}\to\mathcal{B}_bP$ }

\author{Jia-Jie Han$^{1,2}$}       \email{hanjiajie1020@163.com}
\author{Rui-Xiang Zhang$^{1}$}  \email{854474645@qq.com}
\author{Hua-Yu Jiang$^{2,3}$}  \email{jianghy15@lzu.edu.cn}
\author{Zhen-Jun Xiao$^{1}$ } \email{xiaozhenjun@njnu.edu.cn}
\author{Fu-Sheng Yu$^{2,4,5,6}$}   \email{yufsh@lzu.edu.cn}

\affiliation{$^1$ Department of Physics and Institute of Theoretical Physics, Nanjing Normal University, Nanjing, Jiangsu 210023, China}
\affiliation{$^2$ School of Nuclear Science and Technology, Lanzhou University, Lanzhou 730000, China}
\affiliation{$^3$ Theoretische Physik 1, Naturwissenschaftlich-Technische Fakult\"at,
Universit\"at Siegen, 57068 Siegen, Germany}
\affiliation{$^4$ Lanzhou Center for Theoretical Physics, Lanzhou University, Lanzhou, Gansu 730000, China}
\affiliation{$^5$ Frontier Science Center for Rare Isotopes, Lanzhou University,  Lanzhou 730000, China}
\affiliation{$^6$ Center for High Energy Physics, Peking University, Beijing 100871, China}
\date{\today}
\begin{abstract}
After the discovery of the double-charm baryon $\Xi_{cc}^{++}$ by LHCb, one of the most important topics is to search for the bottom-charm baryons which contain a $b$ quark, a $c$ quark and a light quark. In this work, we study the two-body non-leptonic weak decays of  a bottom-charm baryon into a spin-$1/2$ bottomed baryon and a light pseudoscalar meson with the short-distance contributions calculated under the factorization hypothesis and the long-distance contributions considering the final-state-interaction effects. The branching fractions of all  fifty-seven decay channels are estimated. The results indicate that $\Xi_{bc}^+\to\Xi_b^0\pi^+$,  $\Xi_{bc}^{0}\to\Xi_{b}^{-}\pi^+$ and $\Omega_{bc}^0\to\Omega_b^-\pi^+$ decay modes have relatively large decay rates and thus could be used to experimentally search for the bottom-charm baryons. The topological diagrams and the SU(3) symmetry of bottom-charm baryon decays are discussed.
\end{abstract}
\vspace{1cm}
\maketitle

\section{Introduction}\label{sec:intro}

The recent discovery of the double-charm baryon $\Xi_{cc}^{++}$ by the LHCb collaboration \cite{Aaij:2017ueg,Aaij:2018gfl} is the first observation of the 
double-heavy-flavor baryons which contain two heavy-flavor quarks ($b$ or $c$) and one light quark ($u$, $d$ or $s$) \cite{DeRujula:1975qlm,Jaffe:1975us,Ponce:1978gk}. 
It opens a new window to understand the perturbative and non-perturbative features of the strong interaction, since their structure of two heavy quarks in the core and 
one light quark in the cloud is different from all the observed hadrons. It is natural and important to extend our studies to search for the bottom-charm baryons 
($\mathcal{B}_{bc}=\Xi_{bc}^+$, $\Xi_{bc}^0$ or $\Omega_{bc}^0$). The heavy diquark core of $bc$ could be somewhere different from that of $cc$.

The theoretical predictions on the branching fractions of weak decays of the bottom-charm baryons are very helpful for the experimental searches, 
analogous to the case of the observation of the double-charm baryon. There are too many decaying modes of heavy flavor hadrons. The experimental 
measurements prefer the modes with the largest branching fractions and easily detected final particles. In \cite{Yu:2017zst}, it is suggested to search 
for the doubly charmed baryons via $\Xi_{cc}^{++}\to\Lambda_c^+K^-\pi^+\pi^+$ and $\Xi_{cc}^{++}\to\Xi_c^+\pi^+$. 
Subsequently, the LHCb collaboration successfully observed $\Xi_{cc}^{++}$ through the above suggested processes \cite{Aaij:2017ueg,Aaij:2018gfl}. 
The importance of theoretical studies on the weak decays is discussed in \cite{Yu:2019lxw,Xicc2021}.

The charm-quark decays dominate the decay rates of the bottom-charm baryons, like the $B_c$ meson. It is known that around $70\%$ decay rates of $B_c$ 
comes from charm decays, $20\%$ from $b$ quark decays and $10\%$ from $b\bar{c}$ annihilation. The large fractions from charm are induced by the dominate 
weak transitions of $c\to s$ with $|V_{cs}|\sim 1$, while the bottom decays are suppressed by $b\to c$ with $|V_{cb}|\sim 4\times 10^{-2}$. Therefore, in this work, 
we will study the charm decays of the $\mathcal{B}_{bc}$ baryons in order to point out the most favorable modes for experimental searches.

The dynamics of the non-leptonic charm decays are usually difficult to be calculated due to the large non-perturbative contributions at the charm 
scale \cite{Cheng:2010ry,Li:2012cfa,Li:2013xsa}. In the recent years, the weak decays of charmed baryon decays have been extensively studied 
\cite{Lu:2016ogy,Geng:2017esc,Geng:2017mxn,Wang:2017gxe,Geng:2018plk,Cheng:2018hwl,Jiang:2018iqa,Zhao:2018zcb,Geng:2018bow,Geng:2018upx,
He:2018joe,Zhao:2018mov,Grossman:2018ptn,Geng:2018rse,Wang:2019dls,Geng:2019xbo,Hsiao:2019yur,Geng:2019awr,Jia:2019zxi,Zou:2019kzq,
Geng:2020zgr,Niu:2020gjw,Hsiao:2020iwc,Pan:2020qqo,Meng:2020euv,Hu:2020nkg,Cheng:2018rkz}, as well as the weak decays of double-charm or bottom-charm baryons 
\cite{Cheng:2020wmk,Egolf:2002nk,Wang:2017azm,Shi:2017dto,Onishchenko:2000yp,Wang:2017mqp,Jiang:2018oak,Zhang:2018llc,Li:2020qrh,Cheng:2019sxr,Cheng:2018mwu,
Li:2017ndo,Jiang:2018oak,Wang:2017mqp,Wang:2017azm,Shi:2019hbf,Gutsche:2017hux,Hu:2017dzi,Shi:2017dto,Zhao:2018mrg,Xing:2018lre,Dhir:2018twm,Zhang:2018llc,
Shi:2019fph,Xing:2019hjg,Gutsche:2019wgu,Hu:2019bqj,Gutsche:2019iac,Ke:2019lcf,Cheng:2020wmk,Hu:2020mxk,Shi:2020qde,Ivanov:2020xmw,Li:2020qrh}.
In \cite{Yu:2017zst}, the non-leptonic decay amplitudes of the double-charm baryons are calculated with the factorizable contributions using the factorization 
approach and the non-factorizable contribution considering the rescattering mechanism of the final-state-interaction effects. The observation of $\Xi_{cc}^{++}$ 
via the processes predicted by the above theoretical framework manifests that it is reliable for charm decays of double-heavy baryons. This method has been systematically 
used to study the decays of double-charm baryon ($\mathcal{B}_{cc}$) into one charmed baryon and one pseudoscalar meson ($\mathcal{B}_{c} P$) \cite{Xicc2021}, 
one charmed baryon and one vector meson ($\mathcal{B}_{c} V$) \cite{Jiang:2018oak}, and one light octet baryon and one charmed meson ($\mathcal{B} D^{(*)}$) \cite{Li:2020qrh}. 
In this work, we will use the same method to study the bottom-charm baryon decaying into one $b$-baryon and one light pseudoscalar meson, $\mathcal{B}_{bc}\to \mathcal{B}_b P$.

This paper is organized as follows. In section \ref{sec:frame}, we describe theoretical framework adopted in this work. The numerical results and some discussions are collected in 
section \ref{sec:discuss}. Summary is in section \ref{sec:summary}.

\section{Theoretical framework}\label{sec:frame}

The effective Hamiltonian describes the tree-level charm decays is given by:
\begin{equation}
	\mathcal{H}_{eff}=\frac{G_F}{\sqrt{2}}\sum_{q^{\prime}=d,s}^{}V_{cq^{\prime}}^\ast V_{uq}[C_1(\mu)O_1(\mu)+C_2(\mu)O_2(\mu)]+h.c.,
\end{equation}
where $V_{cq^{\prime}}^\ast$ and $V_{uq}$ are CKM matrix elements, $O_1$ and $O_2$ are two four-fermion local operators:
\begin{equation}
	O_1=(\bar{u}_\alpha q_\beta)_{V-A}(\bar{q^{\prime}}_\beta c_\alpha)_{V-A},\quad
	O_2=(\bar{u}_\alpha q_\alpha)_{V-A}(\bar{q^{\prime}}_\beta c_\beta)_{V-A},
\end{equation}
with $C_{1,2}$ as the relevant Wilson coefficients.

\begin{figure}[tp]
	\begin{center}
		\begin{minipage}{0.25\linewidth}	\includegraphics[scale=0.4]{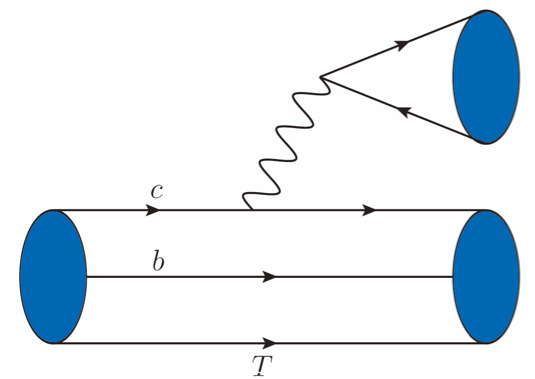}
		\includegraphics[scale=0.4]{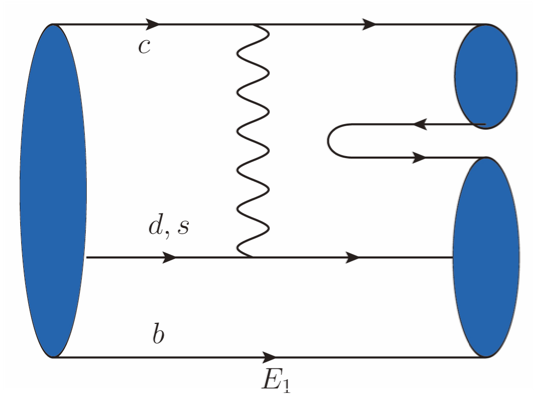}
		\end{minipage}
		\begin{minipage}{0.25\linewidth}	\includegraphics[scale=0.4]{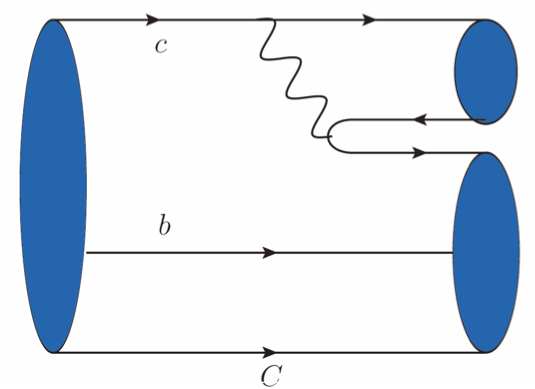} 	\includegraphics[scale=0.4]{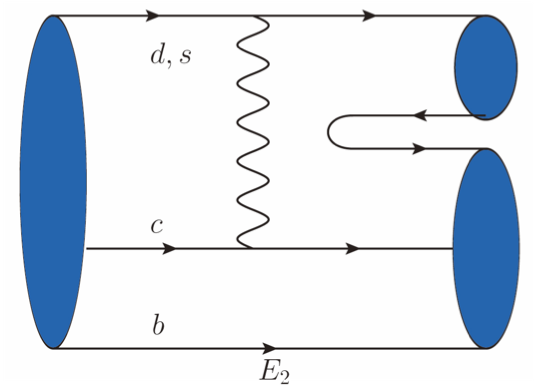}
		\end{minipage}
		\begin{minipage}{0.25\linewidth} 	\includegraphics[scale=0.4]{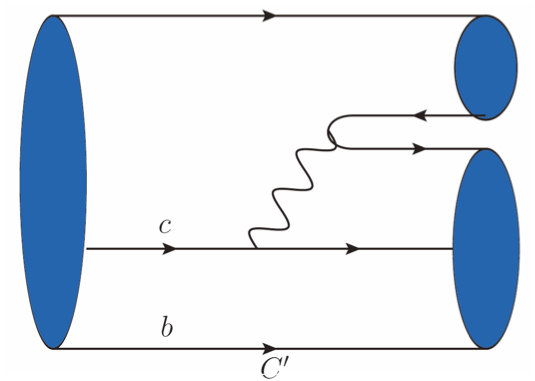}	\includegraphics[scale=0.4]{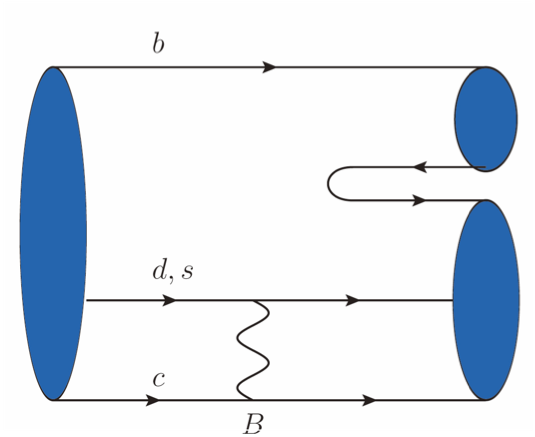}
		\end{minipage}
	\end{center}
	\caption{Tree level topological diagrams for two body nonleptonic decays $\mathcal{B}_{bc}\to\mathcal{B}_b P$  of the bottom-charm baryons 
${\cal B}_{bc} =(\Xi_{bc}^{+, 0}, \Omega_{bc}^0)$. }
	\label{fig:fig1}
\end{figure}

Comparing with the charm-quark decay of $B_c$ meson, the weak decay of $bc$ baryons can receive more complicated contributions even at the tree level. 
These contributions can be divided into several topological diagrams which are depicted in Fig.\ref{fig:fig1}. $T$ denotes color-favored $W$-external emission diagram. 
Different from the weak decay of meson which only receive one color-suppressed contribution and one $W$-exchange contribution, the baryon decays have two 
color-suppressed diagrams, denoted by $C$ and $C^\prime$ in Fig.\ref{fig:fig1}, and three $W$-exchange diagrams described by $E_1$, $E_2$ and $B$. 
The difference between these topological diagrams is from the source of the quarks of the final-state hadrons. For example, the two quarks of the meson 
of the $C$ diagram are both from the weak vertex, while the meson in the $C^\prime$ diagram is formed by one spectator quark and one quark from the 
weak vertex. With the analysis on the hierarchy, the topological diagram in charmed baryon decays are almost at the same order \cite{Leibovich:2003tw,Mantry:2003uz}. 
In the following, we will introduce our framework how to calculate each topological diagrams, for the short-distance and long-distance contributions, respectively.

\subsection{Short-distance contributions under the factorization hypothesis}

In the calculation of the short-distance contributions, $T$-topological diagram is considered as the dominating contribution which can be calculated under the factorization hypothesis. 
The short-distance contributions of the $C$ diagram is negligible due to its smallness compared to $T$ diagram.

The amplitude of $\mathcal{B}_{bc}\to\mathcal{B}_bP$ can be evaluated using the hadronic transition matrix elements:
\begin{equation}
\langle\mathcal{B}_bP|\mathcal{H}_{eff}|\mathcal{B}_{bc}\rangle=\frac{G_F}{\sqrt{2}}V^{\ast}_{cq^{\prime}}V_{uq}\sum_{i=1,2}C_i\langle\mathcal{B}_bP|O_i|\mathcal{B}_{bc}\rangle.
\end{equation}
According to the factorization hypothesis, the matrix elements $\langle\mathcal{B}_bP|O_i|\mathcal{B}_{bc}\rangle$ can be factorized into a product of two parts, 
the decay constants and the heavy-light transition form factors. The color-favored short-distance topological amplitude of  the considered  $\bbc \to \bb P$ weak decays can be expressed as:
\beq
\langle\mathcal{B}_{b}M|\mathcal{H}_{eff}|\mathcal{B}_{bc}\rangle_{SD}^T=\frac{G_F}{\sqrt{2}}V^\ast_{cq^\prime}V_{uq}a_1(\mu)\langle M|\bar{u}\gamma^\mu(1-\gamma_5)q|
0\rangle\langle\mathcal{B}_b|\bar{q}^\prime\gamma_\mu(1-\gamma_5)c|\mathcal{B}_{bc} \rangle . \label{eq:hadronic1}
\eeq
Here, we also give the color-suppressed topology amplitude for convenience:
\beq
\langle\mathcal{B}_{b}M|\mathcal{H}_{eff}|\mathcal{B}_{bc}\rangle_{SD}^C=\frac{G_F}{\sqrt{2}}V^\ast_{cq^\prime}V_{uq}a_2(\mu)\langle M|\bar{q}^\prime\gamma^\mu(1-\gamma_5)q|0\rangle\langle\mathcal{B}_b|\bar{u}\gamma_\mu(1-\gamma_5)c|\mathcal{B}_{bc}\rangle, \label{eq:hadronic2}
\eeq
where $a_1(a_2)$ represents the effective Wilson coefficients, and $a_1(\mu)=C_1(\mu)+C_2(\mu)/3$ for the color-favored
and $a_2(\mu)=C_2(\mu)+C_1(\mu)/3$ for the color-suppressed tree $W$-emission amplitudes, with the Wilson coefficients
$C_1(\mu)=1.21$ and $C_2(\mu)=-0.42$ at the typical scale of charm decays $\mu=m_c=1.3GeV$\cite{Li:2012cfa}.

In Eqs.(\ref{eq:hadronic1}) and (\ref{eq:hadronic2}), the last matrix elements can be parameterized into six form factors:
\begin{equation}
\begin{aligned}
\langle {\cal B}_b(p^\prime,s_z^\prime)| q^{\prime}\gamma_\mu(1-\gamma_5)c |{\cal B}_{bc}(p,s_z)\rangle&=\\
\bar u(p^\prime,s^\prime_z)\Big [ \gamma_\mu f_1(q^2) &+ i\sigma_{\mu\nu}\frac{q^\nu}{M_{\mathcal{B}_{bc}}} f_2(q^2)+\frac{q^\mu}{M_{\mathcal{B}_{bc}}} f_3(q^2) \Big ] u(p,s_z)\\
-\bar u(p^\prime,s^\prime_z)\Big [ \gamma_\mu g_1(q^2) &+ i\sigma_{\mu\nu}\frac{q^\nu}{M_{\mathcal{B}_{bc}}}g_2(q^2) +\frac{q^\mu}{M_{\mathcal{B}_{bc}}} g_3(q^2) \Big ] \gamma_5 u(p,s_z),
\label{eq:ff}
\end{aligned}
\end{equation}
with $q=p-p^\prime$ and $M_{\mathcal{B}_{bc}}$ as the mass of ${\cal B}_{bc}$. The form factors $f_i$ and $g_i$ can be evaluated under different theoretical models.

The first matrix elements in Eq.(\ref{eq:hadronic1}) and (\ref{eq:hadronic2}) are defined as decay constants:
\begin{align}
\langle P(p)|\bar{u}\gamma^\mu(1-\gamma_5)q|0\rangle&=-if_Pp^\mu,\label{eq:Pdecay}\\
\langle V(p)|\bar{u}\gamma^\mu(1-\gamma_5)q|0\rangle&=m_Vf_V\epsilon^{\ast \mu},\label{eq:Vdecay}
\end{align}
where $P$ is a pseudoscalar meson, and V a vector meson, and $\epsilon^\mu$
represents the polarization vector of the vector meson.
Combining Eqs.(\ref{eq:hadronic1}-\ref{eq:Vdecay}),
the short-distance factorizable amplitudes of ${\cal B}_{bc}\to{\cal B}_b P$ are:
\begin{align}
\mathcal{A}_{\mbox{{\tiny SD}}}({\cal B}_{bc}\to{\cal B}_b P)=&i\bar{u}_{\mathcal{B}_b}(A+B\gamma_5)u_{\mathcal{B}_{bc}},\\
\mathcal{A}_{\mbox{{\tiny SD}}}({\cal B}_{bc}\to{\cal B}_b V)=&\epsilon^{\ast \mu}\bar{u}_{\mathcal{B}_b}
\left [ A_1\gamma_\mu\gamma_5
+A_2\frac{p_\mu(\mathcal{B}_b)}{M_{\mathcal{B}_{bc}}}\gamma_5+B_1\gamma_\mu+B_2\frac{p_\mu(\mathcal{B}_b)}{M_{\mathcal{B}_{bc}}}
\right ] u_{\mathcal{B}_{bc}},
\label{eq:B2BP}
\end{align}
where the parameters $A,B$ and $A_{1,2},B_{1,2}$ usually include the whole information of strong interaction. In the factorization
hypothesis, these parameters are expressed as
\begin{align}
A=&\lambda f_P(M_{\mathcal{B}_{bc}}-M_{\mathcal{B}_{b}})f_1(m^2), \hspace{3cm}   B=\lambda f_P(M_{\mathcal{B}_{bc}}+M_{\mathcal{B}_{b}})g_1(m^2),\\
A_1=&-\lambda f_Vm \left ( g_1(m^2)+g_2(m^2)\frac{M_{\mathcal{B}_{bc}}-M_{\mathcal{B}_{b}}}{M_{\mathcal{B}_{bc}}} \right ),
\ \ \ \  A_2=-2\lambda f_Vmg_2(m^2), \\
B_1=&\lambda f_V m \left ( f_1(m^2)-f_2(m^2)\frac{M_{\mathcal{B}_{bc}}+M_{\mathcal{B}_{b}}}{M_{\mathcal{B}_{bc}}} \right ),
\ \ \ \ \ \ \ \ B_2=2\lambda f_Vmf_2(m^2),
\end{align}
where $\lambda=\frac{G_F}{\sqrt{2}}V_{CKM}a_{1,2}(\mu)$, $m$ is the mass of pseudoscalar or vector meson.
$V_{CKM}$ stands for the product of the two relevant CKM matrix elements.

\subsection{Long-distance contributions from the rescattering mechanism}

It is well known that the long-distance contributions are large in the charm decays. The most important problem in charmed hadron weak decays is to calculate the long-distance contributions. The $C$ diagram is dominated by the long-distance contribution, since its short-distance contribution is negligible. The $C'$, $E_1$, $E_2$ and $B$ diagrams are absolutely contributed by the long-distance effects. We will calculate all these topological diagrams.
As stated in the Introduction, we consider the rescattering mechanism of the FSI effects in this work.

\begin{figure}[tb]
	\begin{center}
		\begin{minipage}{0.30\linewidth} 		 \includegraphics[scale=0.65]{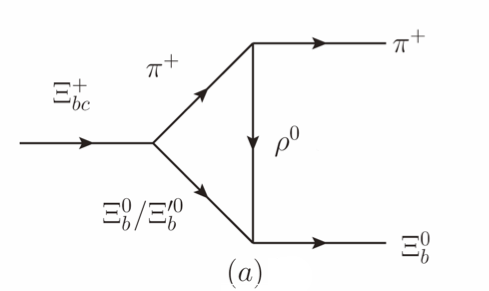}					\includegraphics[scale=0.65]{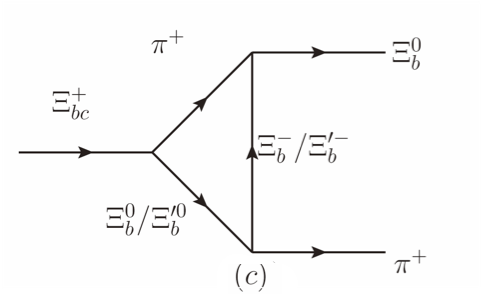} \end{minipage}
		\begin{minipage}{0.30\linewidth} \includegraphics[scale=0.65]{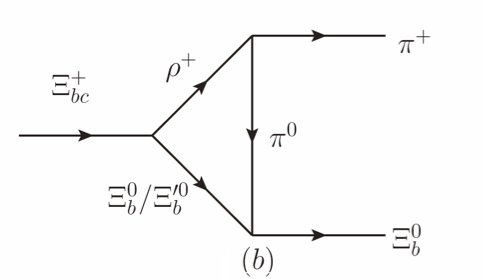}				\includegraphics[scale=0.65]{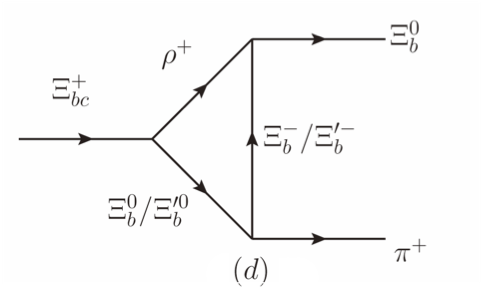} \end{minipage}
	\end{center}
	\caption{The long-distance rescattering contributions to $\Xi_{bc}^{+}\to\Xi_b^0\pi^+$ at hadron level.}
	\label{fig:fig2}
\end{figure}

The final-state interactions can be modeled as the rescattering of two intermediate particles at the hadronic level and described by the single-particle-exchange rescattering diagrams under the hadron-level strong effective Lagrangian. Taking the decay of $\Xi_{bc}^{+}\to\Xi_b^0\pi^+$ as an example, the long-distance contributions come from all the $\Xi_{bc}^+\to(\Xi_b^0/\Xi_b^{\prime 0})\pi^+\to\Xi_b^0\pi^+$ and $\Xi_{bc}^+\to(\Xi_b^0/\Xi_b^{\prime 0})\rho^+\to\Xi_b^0\pi^+$ processes, as shown in Fig.\ref{fig:fig2}. At present consideration, the intermediate states include the light pseudoscalar meson, vector mesons, and
the ground anti-triplet and sextet bottom baryons.

The first weak transition vertex only involves the short-distance contributions calculated in the factorization hypothesis, avoiding the double-counting problem. 
The following scattering process can be evaluated with hadronic strong effective Lagrangian. The theoretical framework is explained in detail in \cite{Xicc2021}.

Now we will take the diagram Fig.\ref{fig:fig2}(b), $\Xi_{bc}^+\to(\Xi_b^0/\Xi_b^{\prime 0})\rho^+\to\Xi_b^0\pi^+$ with a $\pi^0$ exchanged, as a 
concrete example to explain calculation of this triangle diagrams. We adopt the optical theorem and Cutkosky cutting rule as in Ref.\cite{Cheng:FSIB}. The absorptive part of this decay amplitude is calculated as
\begin{align}
{\cal A}bs [\mathcal{M}(P_{\Xi_{bc}^+}\to P_{\pi^+}P_{\Xi_b^0})] &=\frac{1}{2}\int\frac{\mbox{d}^3 p_{\rho^+}}{(2\pi)^3 2E_1}\int\frac{\mbox{d}^3 p_{\Xi_b^0/\Xi_b^{\prime 0}}}{(2\pi)^3 2E_2}(2\pi)^4
\delta^4(p_{\pi^+} + p_{\Xi_b^0} - p_{\rho^+}-p_{\Xi_b^0/\Xi_b^{\prime 0}})\non
& \cdot M(P_{\Xi_{bc}}\to (\Xi_b^0/\Xi_b^{\prime 0})\rho^+)T^*(\pi^+\Xi_b^0\to (\Xi_b^0/\Xi_b^{\prime 0})\rho^+).
\label{eq:optical}
\end{align}

The amplitudes of the weak vertex $\Xi_{bc}^+\to(\Xi_b^0/\Xi_b^{\prime 0})\rho^+$ are taken from Eq.(\ref{eq:B2BP}). The rescattering amplitudes of $(\Xi_b^0/\Xi_b^{\prime 0})\rho^+\to\Xi_b^0\pi^+$ 
can be simply carried out from the hadronic strong-interaction Lagrangian in Appendix A. Then substituting the two parts into Eq.(\ref{eq:optical}), the absorptive part is written as:
{\footnotesize
\begin{align}
	Abs[\mathcal{M}(\rho^+,\Xi_b^0;\pi^0)]=&\int\frac{|\vec{p_1}|sin\theta d\theta d\phi}{32\pi^2m_{\Xi_{bc}^+}}(-ig_{\rho^+\pi^0\pi^-})ig_{\Xi_b^0\pi^0\Xi_b^0}\bar{u}(p_4,s_4)i\gamma_5(\slashed{p_2}+m_2)\frac{F^2(t,m_k)}{t-m_k^2+im_k\Gamma_k}\nonumber\\
	&\hspace{-3cm}\cdot\left((-2\slashed{p_3}+\frac{2p_3\cdot p_1\slashed{p_1}}{m_1^2})(A_1\gamma_5+B_1)+\frac{-2m_1^2p_3\cdot p_2+2p_3\cdot p_1p_1\cdot p_2}{m_1^2m_{\Xi_{bc}^+}}(A_2\gamma_5+B_2)\right)u(p_i,s_i),\\
	Abs[\mathcal{M}(\rho^+,\Xi_b^{\prime 0};\pi^0)]=&\int\frac{|\vec{p_1}|sin\theta d\theta d\phi}{32\pi^2m_{\Xi_{bc}^+}}(-ig_{\rho^+\pi^0\pi^-})ig_{\Xi_b^{\prime 0}\pi^0\Xi_b^0}\bar{u}(p_4,s_4)i\gamma_5(\slashed{p_2}+m_2)\frac{F^2(t,m_k)}{t-m_k^2+im_k\Gamma_k}\nonumber\\
	&\hspace{-3cm}\cdot\left((-2\slashed{p_3}+\frac{2p_3\cdot p_1\slashed{p_1}}{m_1^2})(A_1\gamma_5+B_1)+\frac{-2m_1^2p_3\cdot p_2+2p_3\cdot p_1p_1\cdot p_2}{m_1^2m_{\Xi_{bc}^+}}(A_2\gamma_5+B_2)\right)u(p_i,s_i),
\end{align}}
where $t=p_{\pi^0}^2$. The subscript $i$ denotes the initial state $\Xi_{bc}^+$, while $1,2$ and $k$ represent 
the intermediate states $\rho^+$, $(\Xi_b^0/\Xi_b^{\prime 0})$ and $\pi^0$, and $3,4$ for the final states $\pi^+$ and $\Xi_b^0$, respectively. 
In our calculation, the exchanged $\pi^0$ is generally off-shell, which means the strong coupling constants in the rescattering process are not exactly correct and need to be modified.
To account for the off-shell effect of $\pi^0$, a form factor $F(t,m_\pi)$ is introduced \cite{Cheng:FSIB} as
\begin{align}
F(t,m_\pi)=\frac{\Lambda^2-m_\pi^2}{\Lambda^2-t}, \label{eq:Ffactor}
\end{align}
normalized to unity at the on-shell point $t=p_k^2=m_\pi^2$. The cutoff $\Lambda$ has the form of
\begin{align}
\Lambda= m_\pi + \eta \Lambda_{\rm QCD},
\end{align}
with $\Lambda_{\rm QCD}=330\text{MeV}$ for charm decays. The phenomenological parameter $\eta$ can not be calculated from the first-principle methods, 
and usually needs to be determined by the experimental data. The final results are usually sensitive to the value of $\eta$.
More discussions about $\eta$ can be found in Section \ref{sec:discuss}.

The dispersive part  can be calculated by the dispersion relation of
\begin{align}\label{eq:dispersion}
{\cal D}is[ A(m_1^2)] =\frac{1}{\pi}\int_s^\infty \frac{{\cal A}bs[A(s^\prime)]}{s^\prime - m_1^2} \mbox{d}s^\prime,
\end{align}
which suffers from large ambiguities \cite{Cheng:FSIB}. {The integration in Eq.(19) depends on the knowledge of $\mathcal{A}bs[A(s')]$ in a very large region from $s$ to infinity, which is unavailable currently. Therefore, it is difficult to evaluate how large the dispersive part is. In charm decays as discussed in \cite{Cheng:FSIB,Yu:2017zst,Xicc2021}, the hadronic triangle diagrams would be dominated by the absorptive part, since the total masses of final states are not far from, if not close to, the initial charmed hadrons, especially for charmed baryon decays. In phenomenology, the non-zero dispersive part are absorbed into the only free parameter $\eta$. As $\eta$ can not be calculated from the first principles, it would be finally determined by the experimental data. Then, such effects as the dispersive parts are all involved in the determination of $\eta$, but expected not to affect $\eta$ too much. Besides, the dispersive part mainly affect the studies of CP violations which are sensitive to the strong phases of amplitudes. In the studies of branching fractions, it can be safely involved into the effects of $\eta$.}

By the same approach mentioned in the above section, the absorptive amplitude of the rest triangle diagrams in Fig.\ref{fig:fig2} can be carried out as:
{\footnotesize
\begin{align}
{\cal A}bs[\mathcal{M}(\pi^+,\Xi_b^{0};\rho^0)] =&\int\frac{|\vec{p_1}|sin\theta d\theta d\phi}{32\pi^2m_{\Xi_{bc}^+}}\; i^2(-ig_{\pi^+\rho^0\pi^-})
\frac{F^2(t,m_{\rho^0})}{t-m_{\rho^0}^2+im_{\rho^0}\Gamma_{\rho^0}}  \bar{u}(p_4,s_4) \non
&\hspace{-2cm}\cdot \left[ f_{1\Xi_b^0\rho^0\Xi_b^0}\left (-\slashed{p_1}-\slashed{p_3}+\frac{k\cdot (p_1+p_3)\slashed{k}}{m_{\rho^0}^2} \right )
+\frac{f_{2\Xi_b^0\rho^0\Xi_b^0}}{2m_{\Xi_b^0}} \left (-\slashed{k}(\slashed{p_1}+\slashed{p_3})+k\cdot (p_1+p_3) \right )\right ]\non
&\hspace{-2cm} \cdot (\slashed{p_2}+m_2)(A+B\gamma_5)u(p_i,s_i),\\
Abs[\mathcal{M}(\pi^+,\Xi_b^{\prime 0};\rho^0)]=&\int\frac{|\vec{p_1}|sin\theta d\theta d\phi}{32\pi^2m_{\Xi_{bc}^+}}i^2(-ig_{\pi^+\rho^0\pi^-})\frac{F^2(t,m_{\rho^0})}{t-m_{\rho^0}^2+im_{\rho^0}\Gamma_{\rho^0}}\nonumber\\
&\hspace{-3cm}\cdot\bar{u}(p_4,s_4)\left(f_{1\Xi_b^{\prime 0}\rho^0\Xi_b^0}(-\slashed{p_1}-\slashed{p_3}+\frac{k\cdot (p_1+p_3)\slashed{k}}{m_{\rho^0}^2})+\frac{f_{2\Xi_b^{\prime 0}\rho^0
\Xi_b^0}}{m_{\Xi_b^0}+m_{\Xi_b^{\prime 0}}}(-\slashed{k}(\slashed{p_1}+\slashed{p_3})+k\cdot (p_1+p_3))\right)\nonumber\\
&\hspace{-3cm}\cdot (\slashed{p_2}+m_2)(A+B\gamma_5)u(p_i,s_i),\\
Abs[\mathcal{M}(\pi^+,\Xi_b^0;\Xi_b^-)]=&\int\frac{|\vec{p_1}|sin\theta d\theta d\phi}{32\pi^2m_{\Xi_{bc}^+}}ig_{\pi^+\Xi_b^-\Xi_b^0}g_{\Xi_b^0\Xi_b^-\pi^-}\bar{u}(p_3,s_3)\gamma_5(\slashed{k}+m_k)\gamma_5(\slashed{p_2}+m_2)\nonumber\\
&\cdot(A+B\gamma_5)u(p_i,s_i)\frac{F^2(t,m_{\Xi_b^-})}{t-m_{\Xi_b^-}^2+im_{\Xi_b^-}\Gamma_{\Xi_b^-}},\\
Abs[\mathcal{M}(\pi^+,\Xi_b^0;\Xi_b^{\prime -})]=&\int\frac{|\vec{p_1}|sin\theta d\theta d\phi}{32\pi^2m_{\Xi_{bc}^+}}ig_{\pi^+\Xi_b^{\prime -}\Xi_b^0}g_{\Xi_b^0\Xi_b^{\prime -}\pi^-}\bar{u}(p_3,s_3)\gamma_5(\slashed{k}+m_k)\gamma_5(\slashed{p_2}+m_2)\nonumber\\
&\cdot	(A+B\gamma_5)u(p_i,s_i)\frac{F^2(t,m_{\Xi_b^{\prime -}})}{t-m_{\Xi_b^{\prime -}}^2+im_{\Xi_b^{\prime -}}\Gamma_{\Xi_b^{\prime -}}},\\
Abs[\mathcal{M}(\pi^+,\Xi_b^{\prime 0};\Xi_b^-)]=&\int\frac{|\vec{p_1}|sin\theta d\theta d\phi}{32\pi^2m_{\Xi_{bc}^+}}ig_{\pi^+\Xi_b^-\Xi_b^0}g_{\Xi_b^{\prime 0}\Xi_b^-\pi^-}\bar{u}(p_3,s_3)\gamma_5(\slashed{k}+m_k)\gamma_5(\slashed{p_2}+m_2)\nonumber\\
&\cdot	(A+B\gamma_5)u(p_i,s_i)\frac{F^2(t,m_{\Xi_b^-})}{t-m_{\Xi_b^-}^2+im_{\Xi_b^-}\Gamma_{\Xi_b^-}},\\
Abs[\mathcal{M}(\pi^+,\Xi_b^{\prime 0};\Xi_b^{\prime -})]=&\int\frac{|\vec{p_1}|sin\theta d\theta d\phi}{32\pi^2m_{\Xi_{bc}^+}}ig_{\pi^+\Xi_b^{\prime -}\Xi_b^0}g_{\Xi_b^{\prime 0}\Xi_b^{\prime -}\pi^-}\bar{u}(p_3,s_3)\gamma_5(\slashed{k}+m_k)\gamma_5(\slashed{p_2}+m_2)\nonumber\\
&\cdot		(A+B\gamma_5)u(p_i,s_i)\frac{F^2(t,m_{\Xi_b^{\prime -}})}{t-m_{\Xi_b^{\prime -}}^2+im_{\Xi_b^{\prime -}}\Gamma_{\Xi_b^{\prime -}}},\\
Abs[\mathcal{M}(\rho^+,\Xi_b^0;\Xi_b^-)]=&\int\frac{|\vec{p_1}|sin\theta d\theta d\phi}{32\pi^2m_{\Xi_{bc}^+}}i^3\bar{u}(p_3,s_3)\left(f_{1\rho^+
\Xi_b^-\Xi_b^0}\gamma_\nu-\frac{if_{2\rho^+\Xi_b^-\Xi_b^0}}{m_k+m_3}\sigma_{\mu\nu}p_1^\mu\right)g_{\Xi_b^0\Xi_b^-\pi^+}, \non
&\cdot (\slashed{k}+m_k) \frac{F^2(t,m_{\Xi_b^-})}{t-m_{\Xi_b^-}^2+im_{\Xi_b^-}\Gamma_{\Xi_b^-}}\gamma_5(-g^{\nu\alpha}+\frac{p_1^\nu p_1^\alpha}{m_1^2})
(\slashed{p_2}+m_2) \non
&\cdot  (A_1\gamma_\alpha\gamma_5+A_2\frac{p_{2\alpha}}{m_{bc}^+}\gamma_5+B_1\gamma_\alpha+B_2\frac{p_{2\alpha}}{m_{bc}^+})u(p_i,s_i),\\
Abs[\mathcal{M}(\rho^+,\Xi_b^0;\Xi_b^{\prime -})]=&\int\frac{|\vec{p_1}|sin\theta d\theta d\phi}{32\pi^2m_{\Xi_{bc}^+}}i^3\bar{u}(p_3,s_3)\left(f_{1\rho^+\Xi_b^{\prime -}\Xi_b^0}\gamma_\nu-\frac{if_{2\rho^+\Xi_b^{\prime -}\Xi_b^0}}{m_k+m_3}\sigma_{\mu\nu}p_1^\mu\right)g_{\Xi_b^0\Xi_b^{\prime -}\pi^+}
\non
&\cdot(\slashed{k}+m_k)\frac{F^2(t,m_{\Xi_b^{\prime -}})}{t-m_{\Xi_b^{\prime -}}^2+im_{\Xi_b^{\prime -}}\Gamma_{\Xi_b^{\prime -}}}\gamma_5(-g^{\nu\alpha}+\frac{p_1^\nu p_1^\alpha}{m_1^2})
(\slashed{p_2}+m_2)\non
&\cdot (A_1\gamma_\alpha\gamma_5+A_2\frac{p_{2\alpha}}{m_{bc}^+}\gamma_5+B_1\gamma_\alpha+B_2\frac{p_{2\alpha}}{m_{bc}^+})u(p_i,s_i),\\
Abs[\mathcal{M}(\rho^+,\Xi_b^{\prime 0};\Xi_b^-)]=&\int\frac{|\vec{p_1}|sin\theta d\theta d\phi}{32\pi^2m_{\Xi_{bc}^+}}i^3\bar{u}(p_3,s_3)
\left(f_{1\rho^+\Xi_b^-\Xi_b^0}\gamma_\nu-\frac{if_{2\rho^+\Xi_b^-\Xi_b^0}}{m_k+m_3}\sigma_{\mu\nu}p_1^\mu\right)g_{\Xi_b^{\prime 0}\Xi_b^-\pi^+}
\non
&\cdot  (\ksl+m_k)\frac{F^2(t,m_{\Xi_b^-})}{t-m_{\Xi_b^-}^2+im_{\Xi_b^-}\Gamma_{\Xi_b^-}}\gamma_5
(-g^{\nu\alpha}+\frac{p_1^\nu p_1^\alpha}{m_1^2})(\psl_2+m_2) \non
&\cdot (A_1\gamma_\alpha\gamma_5+A_2\frac{p_{2\alpha}}{m_{bc}^+}\gamma_5+B_1\gamma_\alpha+B_2\frac{p_{2\alpha}}{m_{bc}^+})u(p_i,s_i),\\
Abs[\mathcal{M}(\rho^+,\Xi_b^{\prime 0};\Xi_b^{\prime -})]=&\int\frac{|\vec{p_1}|sin\theta d\theta d\phi}{32\pi^2m_{\Xi_{bc}^+}}i^3\bar{u}(p_3,s_3)
\left(f_{1\rho^+\Xi_b^{\prime -}\Xi_b^0}\gamma_\nu-\frac{if_{2\rho^+\Xi_b^{\prime -}
		\Xi_b^0}}{m_k+m_3}\sigma_{\mu\nu}p_1^\mu\right)g_{\Xi_b^{\prime 0}\Xi_b^{\prime -}\pi^+} \non
&\cdot (\ksl+m_k) \frac{F^2(t,m_{\Xi_b^{\prime -}})}{t-m_{\Xi_b^{\prime -}}^2+im_{\Xi_b^{\prime -}}\Gamma_{\Xi_b^{\prime -}}}
\gamma_5(-g^{\nu\alpha}+\frac{p_1^\nu p_1^\alpha}{m_1^2})(\psl_2+m_2)\non
&\cdot (A_1\gamma_\alpha\gamma_5+A_2\frac{p_{2\alpha}}{m_{bc}^+}\gamma_5+B_1\gamma_\alpha+B_2\frac{p_{2\alpha}}{m_{bc}^+})u(p_i,s_i).
\end{align}}
Collecting all the amplitudes together, the total amplitude of $\Xi_{bc}\to\Xi_b^0\pi^+$ is expressed as:
\begin{align}
\mathcal{A}(\Xi_{bc}^+\to\Xi_b^0\pi^+)=&\mathcal{T}_{SD}(\Xi_{bc}^+\to\Xi_b^0\pi^+)\non
&\hspace{-2cm} + i Abs\Big[\mathcal{M}(\pi^+,\Xi_b^0;\rho^0)+\mathcal{M}(\pi^+,\Xi_b^{\prime 0};\rho^0)+\mathcal{M}(\rho^+,\Xi_b^0;\pi^0)+\mathcal{M}(\rho^+,\Xi_b^{\prime 0};\pi^0)\nonumber\\
& \hspace{-2cm}+\mathcal{M}(\pi^+,\Xi_b^0;\Xi_b^-)+\mathcal{M}(\pi^+,\Xi_b^0;\Xi_b^{\prime -})+\mathcal{M}(\pi^+,\Xi_b^{\prime 0};\Xi_b^-)+\mathcal{M}(\pi^+,\Xi_b^{\prime 0};\Xi_b^{\prime -})\nonumber\\
&\hspace{-2cm}+\mathcal{M}(\rho^+,\Xi_b^0;\Xi_b^-)+\mathcal{M}(\rho^+,\Xi_b^0;\Xi_b^{\prime -})+\mathcal{M}(\rho^+,\Xi_b^{\prime 0};\Xi_b^-)
+\mathcal{M}(\rho^+,\Xi_b^{\prime 0};\Xi_b^{\prime -})\Big], 
\end{align}
where $\mathcal{T}_{SD}$ stands for the short-distance contributions of topological diagram $T$. The amplitudes of all the other channels for $bc$ baryons decaying have been collected into Appendix \ref{app:amp}.

\section{Numerical results and discussions}\label{sec:discuss}

For the calculation of the branching ratios, we need to know all inputs used in this work.
For lack of experimental data, we use the theoretical results of the masses and lifetimes from Ref.\cite{Yu:2018com} and \cite{Berezhnoy:2018bde} as shown in Table \ref{table:input}.
For the bottom-charm baryons, there are two sets of SU(3) triplets, $\Xi_{bc},\Omega_{bc}$ and $\Xi_{bc}^\prime,\Omega_{bc}^\prime$.
Those two triplets have different $J^P$ for the heavy di-quark system $bc$.
Only the lighter ones can weak decay with sizable branching fractions.
The theoretical results obtained in Ref.\cite{Yu:2018com} show that bc-baryons with $J^P=1^+$ are the lighter ones.

The lifetimes are an important feature to calculate the branching fractions and in the experimental measurements \cite{Yu:2019lxw}. They are also calculated in \cite{Cheng:2019sxr} that 409fs$<\tau(\Xi_{bc}^+)<$607fs, 93fs$<\tau(\Xi_{bc}^0)<$118fs, 168fs$<\tau(\Omega_{bc}^0)<$370fs. Compared to those in \cite{Berezhnoy:2018bde}, it can be found that the current status of predictions on the lifetimes of bottom-charm baryons are of large ambiguity, due to the special role of charm. It has to keep in mind that the final results of branching fractions are proportional to the lifetimes. Our following results could be naively replaced by some other values of lifetimes.

\begin{table}[tb]
	\centering
	\caption{Masses (in units of GeV) and lifetime (in units of fs) of the bottom-charmed baryons considered in this paper,
		as given in Refs.~\cite{Yu:2018com,Berezhnoy:2018bde}. }\label{table:input}
	\begin{tabular}{cccc} 	\hline 				\hline
		baryons&$\Xi_{bc}^+$&$\Xi_{bc}^0$&$\Omega_{bc}^0$\\ 				\hline
		masses&6.94&6.94&7.02\\ 				\hline
		lifetime&240&220&180\\
		\hline 	\hline
	\end{tabular}
\end{table}

The masses  and the decay constants of the final-state particles are from \cite{PDG,Choi:2015ywa,Feldmann:1998vh}. The transition form factors in Eq.(\ref{eq:ff}) have been computed out by several methods. We will use the results calculated by the light-front quark model\cite{Wang:2017mqp} as inputs, which have been successfully used in the prediction of the discovery channel $\Xi_{cc}^{++}\to\Lambda_c^+K^-\pi^+\pi^+$ and $\Xi_c^+\pi^+$ in \cite{Yu:2017zst}.
Strong coupling constants can be found in the literatures \cite{Aliev:2010yx,Yan:1992gz,Casalbuoni:1996pg,Meissner:1987ge,Li:2012bt,Aliev:2010nh}, and the unfound ones are calculated with respect to the SU(3) symmetry.
The strong coupling constants appeared in our calculation are gathered in the Appendix \ref{app:lag}.

The branching ratios of  the decay modes $\mathcal{B}_{bc}\to\mathcal{B}_bP$ are listed  in Tables \ref{result1}, \ref{result2}, \ref{result3} and \ref{result4}.
According to the characteristic of the relevant CKM matrix elements for a given decay mode,  all decay modes can be classified into 4 classes:
the short-distance contribution dominated processes in Table \ref{result1}, and the long-distance contribution dominated processes of
the Cabibbo-favored (CF) decays induced by $c\to su\bar{d}$ in Table \ref{result2},
the  singly Cabibbo-suppressed (SCS) ones induced by $c\to du\bar{d}$ or $c\to su\bar{s}$ in Table \ref{result3},
and the doubly Cabibbo-suppressed (DCS) ones induced by $c\to du\bar{s}$  in Table \ref{result4}.
In those tables, the topological amplitudes of each decay mode are listed at the third column with $\lambda_{sd}=V_{cs}^\ast V_{ud}$, $\lambda_{ds}=V_{cd}^\ast V_{us}$, $\lambda_d=V_{cd}^\ast V_{ud}$ and $\lambda_s=V_{cs}^\ast V_{us}$. The {\it tilde} is used to distinguish the channels with sextet bottom-baryons in the final state, from the ones without tilde for the anti-triplet bottom baryons. The branching fractions that involve the long-distance contributions corresponding to three different values of $\eta$, i.e. $\eta$=1.0, 1.5 and 2.0,
are listed at the fifth, sixth and seventh column, respectively.

\begin{figure}[t]
	\centering
	\subfigure{\includegraphics[scale=0.5]{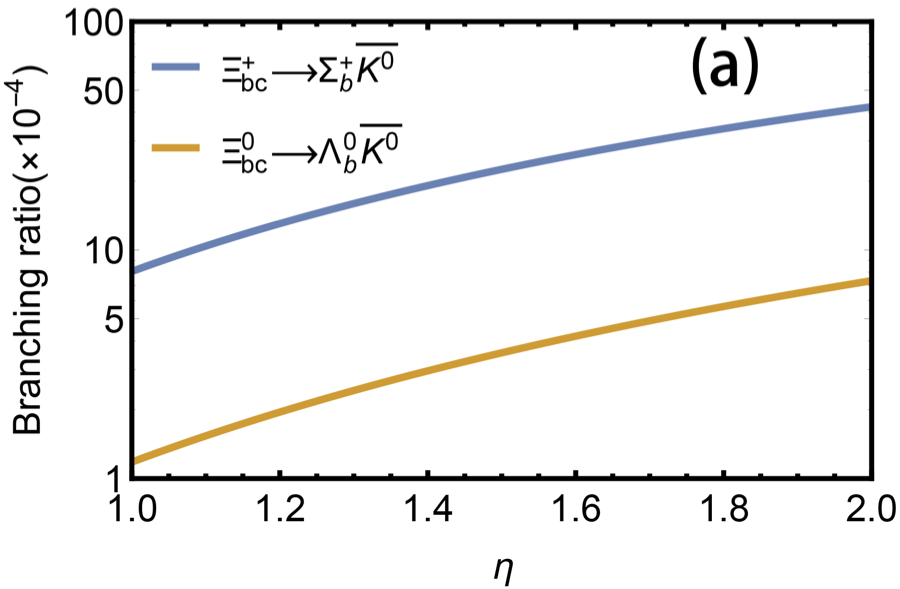}}\hspace{1cm}
	\subfigure{\includegraphics[scale=0.5]{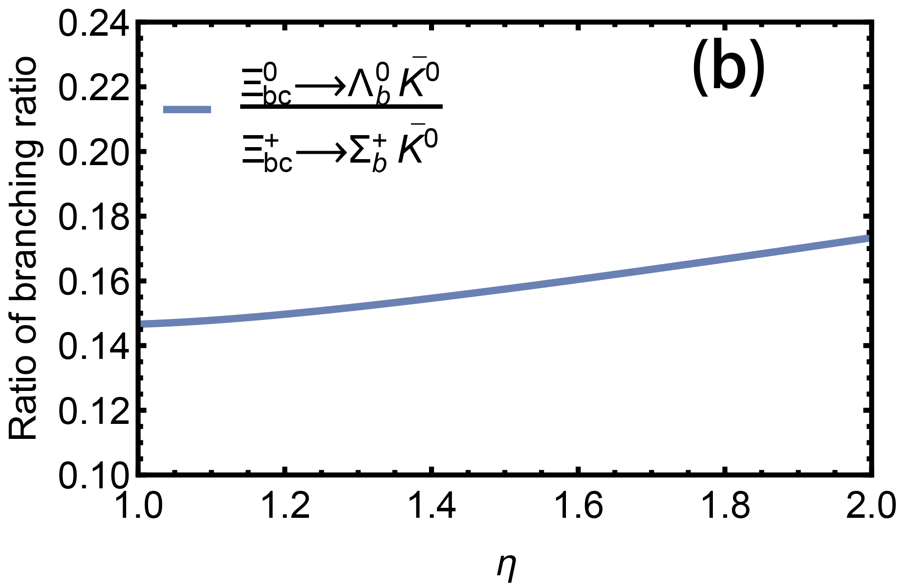}}
	\caption{(a):The theoretical predictions for the branching ratios of $\Xi_{bc}^+\to\Sigma_b^+\bar{K^0}$ and $\Xi_{bc}^0\to\Lambda_b^0\bar{K^0}$ in logarithmic coordinates. (b):Ratio of branching fractions with $\eta$ varying from 1.0 to 2.0.}
	\label{fig:eta}
\end{figure}

The sensitivity of branching ratios to  the parameter $\eta$ in Eq.(\ref{eq:Ffactor}) has already been demonstrated in Refs.~\cite{Yu:2017zst,Jiang:2018oak}.
Taking the decay channels of $\bbc^+ \to \Sigma_b^- \bar{K}^0$ and $\bbc^0 \to \Lambda_b^0 \bar{K}^0$  as an example shown in Fig.~\ref{fig:eta}(a), we can find that the branching fractions can be changed by almost one order of magnitude for $\eta$ varying from $1$ to $2$.
As pointed out in \cite{Yu:2017zst,Xicc2021}, however, the ratios of branching fractions are insensitive to $\eta$, seen in  Fig.~\ref{fig:eta}(b).
Apparently,  the ratio is changed very slightly with the variation of $\eta$. The large uncertainty from the $\eta$ is mostly canceled in the ratio of branching fractions. This behavior helps us to improve the prediction power.

\begin{table}[t]{}
	\centering
	\caption{Branching ratios for the short-distance dominated modes. The ``CF", ``SCS" and ``DCS" represent CKM favored, singly CKM suppressed and doubly CKM suppressed processes, respectively.}
	\label{result1}
	\begin{tabular}{ccc|cccc|c}
		\hline
		\hline
		\rule{0pt}{12pt}Particles&Decay modes&Topology&\ \ \ $\mathcal{BR}_{T_{SD}}(\%)$\ \ \ &$\mathcal{BR}_{\eta=1.0}(\%)$&$\mathcal{BR}_{\eta=1.5}(\%)$&$\mathcal{BR}_{\eta=2.0}(\%)$&\ \ \ CKM\ \ \ \\
		\hline
		\hline
		\rule{0pt}{10pt}$\Xi_{bc}^+$&$\to\Xi_b^0\pi^+$&$\lambda_{sd} (T+C^\prime)$&3.50&3.58&3.73&3.95&CF\\
		\hline
		\rule{0pt}{10pt}&$\to\Xi_b^{\prime 0}\pi^+$&${1\over\sqrt{2}}\lambda_{sd}\big(\tilde{T}+\tilde{C}^{\prime}\big)$&2.04&2.07&2.13&2.21&CF\\
		\hline
		\rule{0pt}{10pt}&$\to\Sigma_b^0\pi^+$&${1\over\sqrt{2}}\lambda_{d}\big(\tilde{T}+\tilde{C}^{\prime}\big)$&0.11&0.12&0.12&0.13&SCS\\
		\hline
		\rule{0pt}{10pt}&$\to\Lambda_b^0\pi^+$&$\lambda_d(T+C^{\prime})$&0.21&0.22&0.23&0.25&SCS\\
		\hline
		\rule{0pt}{10pt}&$\to\Xi_b^{\prime 0}K^+$&${1\over\sqrt{2}}\lambda_s \big(\tilde{T}+\tilde{C}^{\prime}\big)$&0.14&0.14&0.15&0.15&SCS\\
		\hline
		\rule{0pt}{10pt}&$\to\Xi_b^{0}K^+$&$\lambda_s(T+C^{\prime})$&0.30&0.31&0.33&0.35&SCS\\
		\hline
		\rule{0pt}{10pt}&$\to\Lambda_b^0K^+$&$\lambda_{ds}(T+C^\prime)$&0.019&0.020&0.020&0.021&DCS\\
		\hline
		\rule{0pt}{10pt}&$\to\Sigma_b^0K^+$&${1\over\sqrt{2}}\lambda_{ds}\big(\tilde{T}+ \tilde{C}^{\prime}\big)$&0.0086&0.0087&0.0090&0.0094&DCS\\
		\hline
		\hline
		\rule{0pt}{10pt}$\Xi_{bc}^0$&$\to\Xi_b^-\pi^+$&$\lambda_{sd}(T-E_2)$&3.15&3.15&3.23&3.32&CF\\
		\hline
		\rule{0pt}{10pt}&$\to\Xi_b^{\prime -}\pi^+$&${1\over \sqrt{2}} \lambda_{sd} \big(\tilde{T}+\tilde{E}_2\big)$&1.84&1.85&1.86&1.89&CF\\
		\hline
		\rule{0pt}{10pt}&$\to\Sigma_b^-\pi^+$&$\lambda_d(\tilde{T}+\tilde{E}_2)$&0.21&0.21&0.21&0.22&SCS\\
		\hline
		\rule{0pt}{10pt}&$\to\Xi_b^{\prime -}K^+$&${1\over\sqrt{2}}\big(\lambda_s \tilde{T} + \lambda_d \tilde{E}_2\big)$ &0.13&0.13&0.13&0.13&SCS\\
		\hline
		\rule{0pt}{10pt}&$\to\Xi_b^{-}K^+$&$\lambda_s T + \lambda_d E_2$ &0.27&0.27&0.28&0.28&SCS\\
		\hline
		\rule{0pt}{10pt}&$\to\Sigma_b^-K^+$&$\lambda_{ds}\tilde{T}$&0.016&&&&DCS\\
		\hline
		\hline
		\rule{0pt}{10pt}$\Omega_{bc}^0$&$\to\Omega_b^{-}\pi^+$&$\lambda_{sd}\tilde{T}$&2.54&&&&CF\\
		\hline
		\rule{0pt}{10pt}&$\to\Xi_b^{-}\pi^+$&$-\lambda_d T - \lambda_s E_2 $ &0.12&0.12&0.12&0.13&SCS\\
		\hline
		\rule{0pt}{10pt}&$\to\Xi_b^{\prime -}\pi^+$&${1\over\sqrt{2}}\big(\lambda_d \tilde{T} + \lambda_s \tilde{E}_2\big)$ &0.071&0.072&0.072&0.073&SCS\\
		\hline
		\rule{0pt}{10pt}&$\to\Omega_b^{-}K^+$&$\lambda_s\big(\tilde{T}+\tilde{E}_2\big)$&0.17&0.17&0.17&0.18&SCS\\
		\hline
		\rule{0pt}{10pt}&$\to\Xi_b^-K^+$&$\lambda_{ds}(-T+E_2)$ &0.011&0.011&0.011&0.011&DCS\\
		\hline
		\rule{0pt}{10pt}&$\to\Xi_b^{\prime -}K^+$&${1\over \sqrt{2}}\lambda_{ds} \big(\tilde{T}+\tilde{E}_2\big)$&0.0052&0.0052&0.0053&0.0054&DCS\\
		\hline
		\hline
	\end{tabular}
\end{table}

One can see in Table \ref{result1} that the short-distance contributions of the external W-emission $T$ diagrams, $T_{SD}$, are the dominant 
contributions to the modes including it. However, the long-distance dynamics dominate the decay modes without the $T$ diagrams, which can be easily read out from Tables \ref{result2} to \ref{result4}.
For comparison, we display the short-distance contribution of the internal W-emission $C$ diagram under the factorization hypothesis, 
$C_{SD}$, at the fourth column in Tables \ref{result2}-\ref{result4}. It is found that $C_{SD}$ are much smaller than triangle diagram contributions.
The reason is that the effective Wilson coefficient $a_2(\mu)$ at the charm mass scale is deeply suppressed, $a_2(m_c)=-0.017$.

Considering the non-factorizable contributions in the color-suppressed tree-emission diagram, $C$, it is convenient to parameterize such effects in the effective Wilson coefficients $a_{1,2}^{\rm eff}(\mu)=C_{1,2}(\mu)+C_{2,1}(\mu)/N_c^{\rm eff}$ \cite{Ali:1998eb}. The values of $a_{1,2}^{\rm eff}(\mu)$ are actually process dependent. In charmed meson decays, it is found that $|a_2^{\rm eff}/a_1^{\rm eff}|\sim 0.6-0.8$ and $\arg(a_2^{\rm eff}/a_1^{\rm eff})\sim 110^{\circ}-170^{\circ}$ which are obtained from the global fits for the topological amplitudes \cite{Cheng:2010ry,Li:2012cfa}, where the uncertainties are well under control due to the precise measurements of branching fractions. In this work, the non-factorizable effects are modeled by the rescattering mechanism whose uncertainties are characterized by the parameter $\eta$. Since there are not available data currently, we could not determine the value of $\eta$ very well. Varying $\eta$ from 1 to 2, the results could be changed by one order of magnitude for the processes dominated by the non-factorizable effects.

\begin{sidewaystable}
	\centering
	\caption{Branching ratios for the long-distance dominated Cabibbo-favored modes. For the channels
		involving internal W-emission contributions, the short-distance factorizable contributions are also listed at the fourth column for comparison.}
	\label{result2}
	\begin{tabular}{ccc|cccc}
		\hline
		\hline
		\rule{0pt}{12pt}Particles&Decay modes&Topology&\ \ \ $\mathcal{BR}_{T_{SD}}(\times 10^{-3})$\ \ \ &$\mathcal{BR}_{\eta=1.0}(\times 10^{-3})$&$\mathcal{BR}_{\eta=1.5}(\times 10^{-3})$&$\mathcal{BR}_{\eta=2.0}(\times 10^{-3})$\\
		\hline
		\hline
		\rule{0pt}{10pt}$\Xi_{bc}^+$&$\to\Sigma_b^+\bar{K}^0$&$\tilde{C}$&$7.88\times10^{-3}$&0.83&2.31&4.34\\
		\hline
		\hline
		\rule{0pt}{10pt}$\Xi_{bc}^0$&$\to\Omega_b^-K^+$&$\tilde{E}_2$&&0.15&0.55&1.01\\
		\hline				\rule{0pt}{10pt}&$\to\Sigma_b^0\bar{K}^0$&${1\over\sqrt{2}}\big(\tilde{C}+\tilde{E}_1\big)$&$7.07\times10^{-3}$&1.05&3.29&6.80\\
		\hline
		\rule{0pt}{10pt}&$\to\Lambda_b^0\bar{K}^0$&$-C+E_1$&$8.42\times10^{-3}$&0.12&0.35&0.73\\
		\hline
		\rule{0pt}{10pt}&$\to\Sigma_b^+K^-$&$\tilde{E}_1$&&0.36&1.28&2.89\\
		\hline
		\rule{0pt}{10pt}&$\to\Xi_b^{\prime 0}\pi^0$&${1\over2}(-\tilde{C}^{\prime}+\tilde{E}_2)$&$4.66\times10^{-3}$&0.24&0.73&1.48\\
		\hline
		\rule{0pt}{10pt}&$\to\Xi_b^{\prime 0}\eta_1$&${1\over\sqrt{6}}(\tilde{C}^{\prime}+\tilde{E}_1+\tilde{E}_2)$&&&&\\
		\hline
		\rule{0pt}{10pt}&$\to\Xi_b^{\prime 0}\eta_8$&${1\over2\sqrt{3}}(\tilde{C}^{\prime}-2\tilde{E}_1+\tilde{E}_2)$&$5.96\times10^{-3}$&0.039&0.132&0.325\\
		\hline
		\rule{0pt}{10pt}&$\to\Xi_b^{0}\pi^0$&$-{1\over\sqrt{2}}(C^{\prime}+E_2)$&$7.96\times10^{-3}$&1.00&2.89&5.62\\
		\hline
		\rule{0pt}{10pt}&$\to\Xi_b^{0}\eta_1$&${1\over\sqrt{3}}(C^{\prime}+E_1-E_2)$&&&&\\
		\hline
		\rule{0pt}{10pt}&$\to\Xi_b^{0}\eta_8$&${1\over\sqrt{6}}(C^{\prime}-2E_1-E_2)$&&0.021&0.083&0.206\\
		\hline
		\hline
		\rule{0pt}{10pt}$\Omega_{bc}^0$&$\to\Xi_b^{\prime 0}\bar{K}^0$&${1\over\sqrt{2}}(\tilde{C}+\tilde{C}^{\prime})$&$6.34\times10^{-3}$&0.86&2.43&4.63\\
		\hline
		\rule{0pt}{10pt}&$\to\Xi_b^{0}\bar{K}^0$&$-C+C^{\prime}$&$7.33\times10^{-3}$&0.36&1.06&2.08\\
		\hline
		\hline
	\end{tabular}
\end{sidewaystable}

In the same CKM mode, decays with $T_{SD}$ contributions tend to have largest branching fractions.
Under our theoretical framework, $T_{SD}$ is absolutely the dominating contribution comparing to the non-factorizable contributions.
As a result, one can see from the Table \ref{result1} that branching ratios receiving $T_{SD}$ contributions are not sensitive to the variation of $\eta$.
However, the picture is totally different for the other types of contributions which increase or decrease rapidly as $\eta$ changes, which
indicates important FSI effects.

Our results can be compared to those in the previous literatures. Branching ratios of two non-leptonic decay modes of $\Xi_{bc}$ baryon are estimated under NRQCD sum rules in Ref. \cite{Kiselev:2001fw}, i.e. $\mathcal{B}r(\Xi_{bc}^+\to\Xi_b^0\pi^+)=7.7\%$ and $\mathcal{B}r(\Xi_{bc}^0\to\Xi_b^-\pi^+)=7.1\%$, which is almost twice as what we predicted. This deviation is mainly caused by the weak transition form factors. The form factors used in our work are $f_1=0.627$ and $g_1=0.167$ for both decay modes, while $f_1\sim g_1\sim 0.9$ are employed in Ref. \cite{Kiselev:2001fw}. With the form factors calculated in the light-front quark model, Ref. \cite{Wang:2017mqp} estimates the branching ratios of the $\mathcal{T}$-topological diagram involving processes in the factorization approach. Their results are very close to ours, since the form factors used in this work are taken from Ref. \cite{Wang:2017mqp}. Besides, the $\mathcal{T}$-topological diagrams are the dominant contributions in this kind of decay modes. Thus the rescattering effects don't affect the total branching fractions very much.

\begin{sidewaystable}
	\centering
	\caption{The same as Table.\ref{result2} but for singly Cabibbo-suppressed modes.}
	\label{result3}
	\begin{tabular}{ccc|cccc}
		\hline
		\hline
		\rule{0pt}{12pt}Particles&Decay modes&Topology&\ \ \ $\mathcal{BR}_{T_{SD}}(\times 10^{-5})$\ \ \ &$\mathcal{BR}_{\eta=1.0}(\times 10^{-5})$&$\mathcal{BR}_{\eta=1.5}(\times 10^{-5})$&$\mathcal{BR}_{\eta=2.0}(\times 10^{-5})$\\
		\hline
		\hline
		\rule{0pt}{10pt}$\Xi_{bc}^+$&$\to\Sigma_b^+\pi^0$&$-{1\over\sqrt{2}}\tilde{C}$ &$4.65\times10^{-2}$&7.95&26.5&55.2\\
		\hline
		\rule{0pt}{10pt}&$\to\Sigma_b^+\eta_1$&${1\over\sqrt{3}}(\lambda_d+\lambda_s) \tilde{C}$&&&&\\
		\hline
		\rule{0pt}{10pt}&$\to\Sigma_b^+\eta_8$&${1\over\sqrt{6}}(\lambda_d-2\lambda_s) \tilde{C}$&$4.10\times10^{-2}$&6.15&19.2&39.8\\
		\hline
		\hline
		\rule{0pt}{10pt}$\Xi_{bc}^0$&$\to\Sigma_b^{0}\pi^0$&${1\over2}\lambda_d\big(-\tilde{C}-\tilde{C}^{\prime}+\tilde{E}_1+\tilde{E}_2\big)$&$5.19\times10^{-2}$&11.5&35.7&72.7\\
		\hline
		\rule{0pt}{10pt}&$\to\Sigma_b^{0}\eta_1$&${1\over\sqrt{6}}\big[\lambda_d\big(\tilde{C}+\tilde{C}^{\prime}+\tilde{E}_1+\tilde{E}_2\big) + \lambda_s\tilde{C}\big]$&&&&\\
		\hline
		\rule{0pt}{10pt}&$\to\Sigma_b^{0}\eta_8$&${1\over2\sqrt{3}}\big[\lambda_d\big(\tilde{C}+\tilde{C}^{\prime}+\tilde{E}_1+\tilde{E}_2\big) - 2\lambda_s\tilde{C}\big]$&$6.56\times10^{-2}$&6.30&20.3&43.5\\
		\hline
		\rule{0pt}{10pt}&$\to\Lambda_b^{0}\pi^0$&${1\over\sqrt{2}}\lambda_d(C-C^{\prime}-E_1-E_2)$&$6.00\times10^{-2}$&9.81&31.6&66.3\\
		\hline
		\rule{0pt}{10pt}&$\to\Lambda_b^{0}\eta_1$&${1\over\sqrt{3}}\big[\lambda_d\big(C^{\prime}-C+E_1-E_2\big) - \lambda_sC\big]$&&&&\\
		\hline
		\rule{0pt}{10pt}&$\to\Lambda_b^{0}\eta_8$&${1\over\sqrt{6}}\big[\lambda_d\big(C^{\prime}-C+E_1-E_2\big) + 2\lambda_sC\big]$ &$7.88\times10^{-2}$&$0.139$&3.70&6.83\\
		\hline
		\rule{0pt}{10pt}&$\to\Sigma_b^{+}\pi^-$&$\lambda_d \tilde{E}_1$ &&2.18&8.01&18.6\\
		\hline
		\rule{0pt}{10pt}&$\to\Xi_b^{\prime 0}K^0$&${1\over\sqrt{2}} \big(\lambda_s\tilde{C}^{\prime}+\lambda_d\tilde{E}_1\big)$ &$3.25\times10^{-2}$&0.66&2.44&5.88\\
		\hline
		\rule{0pt}{10pt}&$\to\Xi_b^{0}K^0$&$\lambda_sC^{\prime}+\lambda_dE_1$&$6.88\times10^{-2}$&2.92&7.7&13.8\\
		\hline
		\hline
		\rule{0pt}{10pt}$\Omega_{bc}^0$&$\to\Sigma_b^0\bar{K}^0$&${1\over\sqrt{2}}\big(\lambda_d \tilde{C}^{\prime} + \lambda_s \tilde{E}_1\big)$&$1.81\times10^{-2}$&0.82&2.60&5.60\\
		\hline
		\rule{0pt}{10pt}&$\to\Lambda_b^0\bar{K}^0$&$\lambda_dC^{\prime}+\lambda_sE_1$&$3.24\times10^{-2}$&0.70&1.85&3.40\\
		\hline
		\rule{0pt}{10pt}&$\to\Sigma_b^+K^-$&$\lambda_s \tilde{E}_1$ &&1.73&5.82&12.7\\
		\hline
		\rule{0pt}{10pt}&$\to\Xi_b^{\prime 0}\pi^0$&${1\over2}\big(-\lambda_d \tilde{C} + \lambda_s \tilde{E}_2\big)$&$1.80\times10^{-2}$&0.88&2.72&5.69\\
		\hline
		\rule{0pt}{10pt}&$\to\Xi_b^{\prime 0}\eta_1$&${1\over\sqrt{6}}\big[\lambda_s\big(\tilde{C}+\tilde{C}^{\prime}+\tilde{E}_1+\tilde{E}_2\big) + \lambda_d\tilde{C}\big]$&&&&\\
		\hline
		\rule{0pt}{10pt}&$\to\Xi_b^{\prime 0}\eta_8$&${1\over2\sqrt{3}}\big[\lambda_s\big(\tilde{E}_2-2\tilde{C}-2\tilde{C}^{\prime}-2\tilde{E}_1\big) +\lambda_d\tilde{C}\big]$&$4.79\times10^{-2}$&8.92&28.0&58.3\\
		\hline
		\rule{0pt}{10pt}&$\to\Xi_b^{0}\pi^0$&${1\over\sqrt{2}}(\lambda_dC-\lambda_sE_2)$ &$3.060\times10^{-2}$&0.84&2.24&4.21\\
		\hline
		\rule{0pt}{10pt}&$\to\Xi_b^{0}\eta_1$&${1\over\sqrt{3}}\big[\lambda_s\big(C^{\prime}-C+E_1-E_2\big) - \lambda_dC\big] $&&&&\\
		\hline
		\rule{0pt}{10pt}&$\to\Xi_b^{0}\eta_8$&${1\over\sqrt{6}}\big[\lambda_s\big(2C-2C^{\prime}+2E_1-E_2\big) -\lambda_dC\big]$&$6.82\times10^{-3}$&4.42&14.3&30.9\\
		\hline
		\hline
	\end{tabular}
\end{sidewaystable}

\begin{sidewaystable}
	\centering
	\caption{The same as Table.\ref{result2} but for doubly Cabibbo-suppressed modes.}
	\label{result4}
	\begin{tabular}{ccc|cccc}
		\hline
		\hline
		\rule{0pt}{12pt}Particles&Decay modes&Topology&\ \ \ $\mathcal{BR}_{T_{SD}}(\times 10^{-6})$\ \ \ &$\mathcal{BR}_{\eta=1.0}(\times 10^{-6})$&$\mathcal{BR}_{\eta=1.5}(\times 10^{-6})$&$\mathcal{BR}_{\eta=2.0}(\times 10^{-6})$\\
		\hline
		\hline
		\rule{0pt}{10pt}$\Xi_{bc}^+$&$\to\Sigma_b^+K^0$&$\tilde{C}$ &$4.10\times10^{-2}$&$2.50$&$7.69$&15.8\\
		\hline
		\hline
		\rule{0pt}{10pt}$\Xi_{bc}^0$&$\to\Sigma_b^0K^0$&${1\over\sqrt{2}}\big(\tilde{C} + \tilde{C}^\prime\big)$&$3.91\times10^{-2}$&$4.72$&$14.6$&29.6\\
		\hline
		\rule{0pt}{10pt}&$\to\Lambda_b^0K^0$&$-C+C^\prime$&$4.65\times10^{-2}$&$2.12$&$6.65$&$13.7$\\
		\hline
		\hline
		\rule{0pt}{10pt}$\Omega_{bc}^0$&$\to\Sigma_b^0\pi^0$&${1\over2}(-\tilde{E}_1+\tilde{E}_2)$&&$0.403$&$1.35$&$2.90$\\
		\hline
		\rule{0pt}{10pt}&$\to\Sigma_b^0\eta_1$&${1\over\sqrt{6}}\big(\tilde{C}^\prime+\tilde{E}_1+\tilde{E}_2\big)$&&&&\\
		\hline
		\rule{0pt}{10pt}&$\to\Sigma_b^0\eta_8$&${1\over2\sqrt{3}}\big(-2\tilde{C}^\prime+\tilde{E}_1+\tilde{E}_2\big)$&$1.46\times10^{-2}$&$0.53$&$1.63$&$3.40$\\
		\hline
		\rule{0pt}{10pt}&$\to\Lambda_b^0\pi^0$&$-{1\over\sqrt{2}}(E_1+E_2)$&&$1.21$&$4.03$&8.72\\
		\hline
		\rule{0pt}{10pt}&$\to\Lambda_b^0\eta_1$&${1\over\sqrt{3}}(C^\prime+E_1-E_2)$&&&&\\
		\hline
		\rule{0pt}{10pt}&$\to\Lambda_b^0\eta_8$&$-{1\over\sqrt{6}}(2C^\prime-E_1+E_2)$&$3.10\times10^{-2}$&$1.67$&$5.00$&10.1\\
		\hline
		\rule{0pt}{10pt}&$\to\Sigma_b^-\pi^+$&$\tilde{E}_2$ &&$0.395$&$1.31$&$2.92$\\
		\hline
		\rule{0pt}{10pt}&$\to\Sigma_b^+\pi^-$&$\tilde{E}_1$ &&$0.99$&$3.51$&$8.03$\\
		\hline
		\rule{0pt}{10pt}&$\to\Xi_b^{\prime 0}K^0$&${1\over\sqrt{2}}\big(\tilde{C}+\tilde{E}_1\big)$&$1.32\times10^{-2}$&$2.55$&$8.59$&18.9\\
		\hline
		\rule{0pt}{10pt}&$\to\Xi_b^{0}K^0$&$-C+E_1$&$2.73\times10^{-2}$&$0.408$&$1.04$&2.01\\
		\hline
		\hline
	\end{tabular}
\end{sidewaystable}
In Ref.~\cite{Yu:2017zst}, the process of $\Xi_{cc}^{++}\to\Xi_c^+\pi^+$ is recommended as a discovery channel of $\Xi_{cc}^{++}$, which has been confirmed by the LHCb experiment in Ref.~\cite{Aaij:2018gfl}.
Similarly here, we suggest the LHCb collaboration to search for the bottom-charm baryons by the modes of $\Xi_{bc}^{+}\to\Xi_b^0\pi^+$, $\Xi_{bc}^{0}\to\Xi_b^-\pi^+$
and $\Omega_{bc}^{0}\to\Omega_b^-\pi^+$, respectively, who have the relatively larger branching ratios.
The branching ratios of those three decays are given as:
\begin{align}
\mathcal{BR}(\Xi_{bc}^{+}\to\Xi_b^0\pi^+)&= (3.50\sim3.95)\%, \\
\mathcal{BR}(\Xi_{bc}^{0}\to\Xi_b^-\pi^+)&= (3.15\sim3.32)\%,\\
\mathcal{BR}(\Omega_{bc}^{0}\to\Omega_b^-\pi^+)&=2.54\%.
\end{align}

With the branching fractions of $\Xi_{bc}^+\to \Sigma_b^+\bar{K^0}$ and $\Xi_{bc}^+\to \Xi_b^{\prime 0}\pi^+$ in Table.\ref{result2} and \ref{result1}, we calculate the ratio between long-distance contribution of topological diagram $C$ and short-distance contribution of $T$ with $\eta=1.0\sim2.0$,
\begin{align}
\frac{|\mathcal{A}(\Xi_{bc}^+\to \Sigma_b^+\bar{K^0})|}{|\mathcal{A}(\Xi_{bc}^+\to \Xi_b^{\prime 0}\pi^+)|}\approx\frac{C_{LD}}{T_{SD}}\approx 0.2\sim 0.44\ \ \ \
\label{eq33}
\end{align}

In Table.\ref{result2}, there are three channels only include one kind of topological diagrams, i.e. $\Xi_{bc}^+\to\Sigma_b^+\bar{K^0}$, $\Xi_{bc}^0
\to\Omega_b^-K^+$ and $\Xi_{bc}^0\to\Sigma_b^+K^-$. They are more convenient for the studies on the topological diagrams. Although, we can extract the absolute value of the corresponding topological diagram
amplitude, the ratios between these amplitudes have a more important sense. Our numerical results are
\begin{align}
\frac{|\mathcal{A}(\Xi_{bc}^0\to\Sigma_b^+K^-)|}{|\mathcal{A}(\Xi_{bc}^+\to\Sigma_b^+\bar{K^0})|}&=\frac{|E_1|}{|C|}=0.69\sim 0.86,\\
\frac{|\mathcal{A}(\Xi_{bc}^0\to\Omega_b^-K^+)|}{|\mathcal{A}(\Xi_{bc}^+\to\Sigma_b^+\bar{K^0})|}&=\frac{|E_2|}{|C|}=0.52\sim 0.59.
\end{align}
We can see that the ratios ${|E_1|/|C|}$ and ${|E_2|/|C|}$ are all at the order one, confirmed to the result in \cite{Leibovich:2003tw}.
Similarly, the ratios can also be obtained from Table.\ref{result3},
\begin{align}
\frac{|\mathcal{A}(\Xi_{bc}^0\to\Sigma_b^+\pi^-)|}{|\mathcal{A}(\Xi_{bc}^+\to\Sigma_b^+\pi^0)|}&=\frac{|E_1|}{|C|}=0.55\sim 0.62,\\
\frac{|\mathcal{A}(\Omega_{bc}^0\to\Sigma_b^+K^-)|}{|\mathcal{A}(\Xi_{bc}^+\to\Sigma_b^+\pi^0)|}&=\frac{|E_1|}{|C|}=0.56\sim 0.58,
\end{align}
and from Table \ref{result4},
\begin{align}
\frac{|\mathcal{A}(\Omega_{bc}^0\to\Sigma_b^-\pi^+)|}{|\mathcal{A}(\Xi_{bc}^+\to\Sigma_b^+K^0)|}&=\frac{|E_2|}{|C|}=0.43\sim 0.46,\\
\frac{|\mathcal{A}(\Omega_{bc}^0\to\Sigma_b^+\pi^-)|}{|\mathcal{A}(\Xi_{bc}^+\to\Sigma_b^+K^0)|}&=\frac{|E_1|}{|C|}=0.68\sim 0.77.
\label{eq39}
\end{align}
It is clear that all the results are consistent within the theoretical uncertainties.

The $C^{\prime}$ diagram can be indirectly extracted from the channel with the topological amplitudes of $T+C^{\prime}$ in Table \ref{result1}.
The ratios between $C'$ and $C$ are given by:
\begin{align}
\frac{|\mathcal{A}(\Xi_{bc}^+\to\Xi_b^0\pi^+)_{LD}|}{|\mathcal{A}(\Xi_{bc}^+\to\Sigma_b^+\bar{K^0})|}=&\frac{|C^{\prime}|}{|C|}=0.91\sim 0.95,\\
\frac{|\mathcal{A}(\Xi_{bc}^+\to\Lambda_b^0\pi^+)_{LD}|}{|\mathcal{A}(\Xi_{bc}^+\to\Sigma_b^+\pi^0)|}=&\frac{|C^{\prime}|}{|C|}=0.82\sim 0.84,\\
\frac{|\mathcal{A}(\Xi_{bc}^+\to\Xi_b^0K^+)_{LD}|}{|\mathcal{A}(\Xi_{bc}^+\to\Sigma_b^+\pi^0)|}=&\frac{|C^{\prime}|}{|C|}=0.99\sim 1.03,\\
\frac{|\mathcal{A}(\Xi_{bc}^+\to\Lambda_b^0K^+)_{LD}|}{|\mathcal{A}(\Xi_{bc}^+\to\Sigma_b^+K^0)|}=&\frac{|C^{\prime}|}{|C|}=0.97\sim 1.0
\label{eq:42}
\end{align}
From Eqs.~(\ref{eq33}-\ref{eq:42}), we get the similar conclusion as in Refs.~\cite{Leibovich:2003tw,Mantry:2003uz}.
\begin{align}
\frac{|C|}{|T|}\sim\frac{|C^\prime|}{|C|}\sim\frac{|E_1|}{|C|}\sim\frac{|E_2|}{|C|}\sim O(\Lambda_{QCD}/m_c)\sim O(1).
\end{align}

In heavy quark decays, the flavor $SU(3)$ symmetry is a useful tool. A number of relations between decay widths are found within the $SU(3)$ symmetry \cite{Wang:2017azm}. To test the $SU(3)$ symmetry and its breaking, 
we show the ratios in our calculations in Eqs.(\ref{eq:SU3-1}-\ref{eq:SU3-9}). All of these ratios should be unity in the $SU(3)$ limit.
The corresponding topological amplitudes are given as well.
{\small
	\begin{align}
		\Gamma\big( \Xi_{bc}^{+}\to\Lambda_b^0\pi^+ \big)/\Gamma\big( \Xi_{bc}^{+}\to\Xi_b^0K^+ \big) & = | \lambda_d(T+C^\prime) |^2 / |\lambda_s(T+C^\prime)|^2 = 0.70\sim 0.71, \label{eq:SU3-1}\\
		\Gamma\big( \Xi_{bc}^{0}\to\Xi_b^0K^0 \big)/ \Gamma\big( \Omega_{bc}^{0}\to\Lambda_b^0\overline{K}^0 \big) &= |\lambda_s C^\prime+\lambda_d E_1|^2/|\lambda_d C^\prime+\lambda_s E_1|^2 = 2.94\sim 3.08, \label{eq:SU3-2}\\
	\Gamma\big( \Omega_{bc}^{0}\to\Xi_b^-\pi^+ \big)/\Gamma\big( \Xi_{bc}^{0}\to\Xi_b^-K^+ \big) & = |-\lambda_d T-\lambda_s E_2|^2/|\lambda_s T+\lambda_d E_2|^2= 0.58\sim 0.59, \label{eq:SU3-3}\\
	\Gamma\big( \Xi_{bc}^{+}\to\Sigma_b^{+}\pi^0 \big)/\frac{1}{3}\Gamma\big( \Xi_{bc}^{+}\to\Sigma_b^{+}\eta_8 \big) &= |-\frac{1}{\sqrt{2}}\tilde{C}|^2/|-\frac{1}{\sqrt{6}}(\lambda_d-2\lambda_s)\tilde{C}|^2= 5.89\sim 6.57, \label{eq:SU3-4}\\
	\Gamma\big( \Xi_{bc}^{+}\to\Sigma_b^{0}\pi^+ \big)/\Gamma\big( \Xi_{bc}^{+}\to\Xi_b^{\prime 0}K^+ \big) &= |\frac{1}{\sqrt{2}}\lambda_d(\tilde{T}+\tilde{C}^\prime)|^2/|
\frac{1}{\sqrt{2}}\lambda_s(\tilde{T}+\tilde{C}^\prime)|^2=0.80\sim 0.81, \label{eq:SU3-5}\\
	\Gamma\big( \Xi_{bc}^{0}\to\Sigma_b^{+}\pi^- \big)/\Gamma\big( \Omega_{bc}^{0}\to\Sigma_b^{+}K^- \big)&=|\lambda_d \tilde{E}_1|^2/|\lambda_s \tilde{E}_1|^2=0.89\sim 1.02, \label{eq:SU3-6}\\
	\Gamma\big( \Xi_{bc}^{0}\to\Sigma_b^{-}\pi^+ \big)/\Gamma\big( \Omega_{bc}^{0}\to\Omega_b^{-}K^+ \big)&=|\lambda_d(\tilde{T}+\tilde{E}_2)|^2/|\lambda_s(\tilde{T}+\tilde{E}_2)|^2= 0.90\sim 0.91, \label{eq:SU3-7}\\
	\Gamma\big( \Xi_{bc}^{0}\to\Xi_b^{\prime 0}K^0 \big)/\Gamma\big( \Omega_{bc}^{0}\to\Sigma_b^{0}\overline{K}^0 \big) &=|\frac{1}{\sqrt{2}}(\lambda_s\tilde{C}^\prime+\lambda_d\tilde{E}_1)|^2/
|\frac{1}{\sqrt{2}}(\lambda_d\tilde{C}^\prime+\lambda_s\tilde{E}_1)|^2= 0.60\sim 0.77,\label{eq:SU3-8}\\
	\Gamma\big( \Omega_{bc}^{0}\to\Xi_b^{\prime -}\pi^+ \big)/\Gamma\big( \Xi_{bc}^{0}\to\Xi_b^{\prime -}K^+ \big)&=|\frac{1}{\sqrt{2}}(\lambda_d\tilde{T}^\prime+\lambda_s\tilde{E}_2)|^2/
|\frac{1}{\sqrt{2}}(\lambda_s\tilde{T}^\prime+\lambda_d\tilde{E}_2)|^2= 0.38\sim 0.72. \label{eq:SU3-9}
	\end{align}}
The above results indicate that the long-distance final-state interactions can contribute to large $SU(3)$ breaking effect. It would be of great help to research the 
$SU(3)$ symmetry in the heavy baryon decays in case that the long-distance dynamics can be calculated accurately.

\section{Summary}\label{sec:summary}

In this work, we investigate the two-body non-leptonic weak decays of the bottom-charm baryons $\Xi_{bc}^+$, $\Xi_{bc}^0$ and $\Omega_{bc}^0$ with the long-distance 
contributions included. This long-distance contributions, which are non-perturbative and can not be calculated under factorization hypothesis, are viewed as the final-state-interactions in this paper. 
The rescattering mechanism has been adopted for the calculations.

This paper exhibit the calculated branching ratios that include all the processes $\mathcal{B}_{bc}\to\mathcal{B}_bP$.
Based on our results of all fifty-seven decay modes considered in this, 
we recommend the three processes $\Xi_{bc}^{+}\to\Xi_b^0\pi^+$, $\Xi_{bc}^{0}\to\Xi_b^-\pi^+$ and $\Omega_{bc}^{0}\to\Omega_b^-\pi^+$ as the potentials of the discovery channels.
The $T$ topological diagrams are found
to have the largest contribution. Short-distance factorizable contribution $T_{SD}$ is dominating if one decay mode can receive $T$-topology contribution and the branching ratio of this type of decay mode is not sensitive to the variation of $\eta$. However, branching ratios with $C_{SD}$ or purely nonfactorizable contributions vary rapidly
as $\eta$ changes while the ratio between any two branching fractions is almost independent of the variation of $\eta$.

We also calculate the ratios between different topoloical diagrams, and find that in charm decay, the relations 
${|C|\over |T|}\sim{|C^\prime|\over |C|}\sim{|E_1|\over |C|}\sim{|E_2|\over |C|}\sim O({\Lambda_{\text{QCD}}\over m_c})\sim O(1)$ 
are hold as declared in  literatures\cite{Leibovich:2003tw,Mantry:2003uz}. 
The SU(3) breaking effects between decay modes are calculated and some larger SU(3) breaking decay modes have been found 
which have significant guidance for the following researches of bottom-charm baryon decays.

\section{Acknowledgement}
This work was supported by the National Natural Science Foundation of China under the Grant No.~11775117,~U1732101,~11975112, 
and National Key Research and Development Program of China under Contracts Nos. 2020YFA0406400.


\appendix

\section{ Effective Lagrangians}\label{app:lag}

The effective Lagrangians used in the rescattering mechanism are those as given in
Refs.~\cite{Aliev:2010yx,Yan:1992gz,Casalbuoni:1996pg,Meissner:1987ge,Li:2012bt}:
\begin{align}
\mathcal{L}_{VPP}=&\frac{ig_{VPP}}{\sqrt{2}}Tr[V^\mu[P,\partial_\mu P] ],\\
\mathcal{L}_{VVV}=&\frac{ig_{VVV}}{\sqrt{2}}Tr[(\partial_\nu V_\mu V^\mu-V^\mu\partial_\nu V_\mu)V^\nu],\\
\mathcal{L}_{PB_6B_6}=&g_{PB_6B_6}Tr[\bar{B}_6i\gamma_5PB_6],\\
\mathcal{L}_{PB_{\bar{3}}B_{\bar{3}}}=&g_{PB_{\bar{3}}B_{\bar{3}}}Tr[\bar{B}_{\bar{3}}i\gamma_5PB_{\bar{3}}],\\
\mathcal{L}_{PB_6B_{\bar{3}}}=&g_{PB_6B_{\bar{3}}}Tr[\bar{B}_6i\gamma_5PB_{\bar{3}}]+h.c.,\\
\mathcal{L}_{VB_6B_6}=&f_{1VB_6B_6}Tr[\bar{B}_6\gamma_\mu V^\mu B_6]+\frac{f_{2PB_6B_6}}{2m_6}Tr[\bar{B}_6\sigma_{\mu\nu}\partial^\mu V^\nu B_6],\\
\mathcal{L}_{VB_{\bar{3}}B_{\bar{3}}}=&f_{1PB_{\bar{3}}B_{\bar{3}}}Tr[\bar{B}_{\bar{3}}\gamma_\mu V^\mu B_{\bar{3}}]
+\frac{f_{2PB_{\bar{3}}B_{\bar{3}}}}{2m_3}Tr[\bar{B}_{\bar{3}}\sigma_{\mu\nu}\partial^\mu V^\nu B_{\bar{3}}],\\		
\mathcal{L}_{VB_6B_{\bar{3}}}=& \{f_{1VB_6B_{\bar{3}}}Tr[\bar{B}_6\gamma_\mu V^\mu B_{\bar{3}}]
+\frac{f_{2VB_6B_{\bar{3}}}}{m_6+m_3}Tr[\bar{B}_6\sigma_{\mu\nu}\partial^\mu V^\nu B_{\bar{3}}]\}+h.c.,
\end{align}
\beq
P(J^P=0^-)=\left(
\begin{array}{ccc}
	\frac{\pi^0}{\sqrt{2}}+\frac{\eta_8}{\sqrt{6}}&\pi^+&K^+\\
	\pi^-&-\frac{\pi^0}{\sqrt{2}}+\frac{\eta_8}{\sqrt{6}}&K^0\\
	K^-&\bar{K}^0&-\sqrt{\frac{2}{3}}\eta_8\\
\end{array}
\right)+\frac{1}{\sqrt{3}}\left(
\begin{array}{ccc}
	\eta_1&0&0\\
	0&\eta_1&0\\
	0&0&\eta_1\\
\end{array}
\right),
\eeq
\beq
V(J^P=1^-) =\left(
\begin{array}{ccc}
	\frac{\rho^0}{\sqrt{2}}+\frac{\omega}{\sqrt{2}}&\rho^+&K^{\ast +}\\
	\rho^-&-\frac{\rho^0}{\sqrt{2}}+\frac{\omega}{\sqrt{2}}&K^{\ast 0}\\ 		K^{\ast -}&\bar{K}^{\ast 0}&\phi\\
\end{array}
\right) ,
\eeq
\beq
B_6(J^P=\frac{1}{2}^+)=\left(
\begin{array}{ccc}
	\Sigma_b^{+}&\frac{\Sigma_b^0}{\sqrt{2}}&\frac{\Xi_b^{\prime 0}}{\sqrt{2}}\\
	\frac{\Sigma_b^0}{\sqrt{2}}&\Sigma_b^-&\frac{\Xi_b^{\prime -}}{\sqrt{2}}\\
	\frac{\Xi_b^{\prime 0}}{\sqrt{2}}&\frac{\Xi_b^{\prime -}}{\sqrt{2}}&\Omega_b^-\\
\end{array}
\right), \quad
B_{\bar{3}}(J^P=\frac{1}{2}^+)=\left(
\begin{array}{ccc}
	0&\Lambda_b^0&\Xi_b^0\\
	-\Lambda_b^0&0&\Xi_b^-\\
	-\Xi_b^0&-\Xi_b^-&0\\
\end{array}\right).
\eeq

Strong coupling constants are collected in Table.\ref{ap:VPP}, \ref{ap:PBB} and \ref{ap:VBB}.

\begin{table}[h]
	\centering
	\caption{Strong coupling constants of VPP and VVV vertices.}
	\label{ap:VPP}
	\begin{tabular}{|c|c|c|c|c|c|c|c|c|c|}
		\hline
		\hline
		\rule{0pt}{12pt}Vertex&g&Vertex&g&Vertex&g&Vertex&g&Vertex&g\\
		\hline
		\hline
		\rule{0pt}{12pt}$\rho^+\to\pi^0\pi^+$&6.05&$\rho^0\to\pi^+\pi^-$&6.05&$\rho^+\to K^+\bar{K^0}$&4.60&$\rho^0\to K^0\bar{K^0}$&-3.25&$\rho^0\to K^+K^-$&3.25\\
		\hline
		\rule{0pt}{12pt}$\phi\to K^-K^+$&4.60&$\bar{K^{\ast 0}}\to\eta_8\bar{K^0}$&5.63&$\bar{K^{\ast 0}}\to K^-\pi^+$&4.60&$\bar{K^{\ast 0}}\to \bar{K^0}\pi^0$&-3.25&$K^{\ast +}\to K^+\pi^0$&3.25\\
		\hline
		\rule{0pt}{12pt}$K^{\ast +}\to\eta_8 K^+$&5.63&$K^{\ast +}\to\pi^+K^0$&4.60&$K^{\ast 0}\to\pi^-K^+$&4.60&$K^{\ast 0}\to K^0\eta_8$&5.63&$K^{\ast 0}\to\pi^0K^0$&-3.25\\
		\hline
		\rule{0pt}{12pt}$\omega\to K^+K^-$&3.25&$\phi\to\bar{K^0}K^0$&4.60&$\omega\to K^0\bar{K^0}$&3.25&&\\
		\hline
		\hline
		\rule{0pt}{12pt}$\rho^+\to\rho^0\rho^+$&7.38&$\rho^0\to\rho^-\rho^+$&7.38&$\rho^+\to K^{\ast +}\bar{K^{\ast 0}}$&5.22&$\rho^0\to K^{\ast +}K^{\ast -}$&3.69&$\omega\to K^{\ast +}K^{\ast -}$&3.69\\
		\hline
		\rule{0pt}{12pt}$\bar{K^{\ast 0}}\to\phi\bar{K^{\ast 0}}$&5.22&$\bar{K^{\ast 0}}\to\bar{K^{\ast 0}}\rho^0$&-3.69&$\bar{K^{\ast 0}}\to\bar{K^{\ast 0}}\omega$&3.69&$K^{\ast +}\to\rho^+K^{\ast 0}$&5.22&$K^{\ast +}\to\phi K^{\ast +}$&5.22\\
		\hline
		\rule{0pt}{12pt}$K^{\ast +}\to\omega K^{\ast +}$&3.69&$K^{\ast 0}\to \rho^0 K^{\ast 0}$&-3.69&$K^{\ast 0}\to \omega K^{\ast 0}$&3.69&$K^{\ast 0}\to K^{\ast 0}\phi$&5.22&$\phi\to K^{\ast -}K^{\ast +}$&5.22\\
		\hline
		\rule{0pt}{12pt}$\omega\to K^{\ast 0}\bar{K^{\ast 0}}$&3.69&$\phi\to\bar{K^{\ast 0}}K^{\ast 0}$&5.22&$\rho^0\to K^{\ast 0}\bar{K^{\ast 0}}$&-3.69&$K^{\ast +}\to K^{\ast +}\rho^0$&3.69&$\bar{K^{\ast 0}}\to K^{\ast -} \rho^+$&5.22\\
		\hline
		\hline
	\end{tabular}
\end{table}

\begin{table}[h]
	\centering
	\caption{Strong coupling constants of $P\mathcal{B}_{\bar{3}}\mathcal{B}_{\bar{3}}$, $P\mathcal{B}_{\bar{3}}\mathcal{B}_6$ and $P\mathcal{B}_6\mathcal{B}_6$ vertex.}
	\label{ap:PBB}
	\begin{tabular}{|c|c|c|c|c|c|c|c|c|c|}
		\hline
		\hline
		\rule{0pt}{12pt}Vertex&g&Vertex&g&Vertex&g&Vertex&g&Vertex&g\\
		\hline
		\hline
		\rule{0pt}{12pt}$\Xi_b^0\to\Lambda_b^0\bar{K^0}$&1.5&$\Lambda_b^0\to\Xi_b^0K^0$&1.5&$\Xi_b^0\to\Xi_b^0\eta_8$&-0.5&$\Lambda_b^0\to\Lambda_b^0\eta_8$&1.0&$\Xi_b^-\to\Lambda_b^0K^-$&-1.5\\
		\hline
		\rule{0pt}{12pt}$\Xi_b^-\to\Xi_b^-\eta_8$&-0.41&$\Xi_b^-\to\Xi_b^0\pi^-$&1.0&$\Xi_b^0\to\Xi_b^-\pi^+$&1.0&$\Xi_b^-\to\Xi_b^-\pi^0$&-0.71&$\Xi_b^0\to\Xi_b^0\pi^0$&1.0\\
		\hline
		\rule{0pt}{12pt}$\Sigma_b^0\to\Xi_b^0 K^0$&-11.5&$\Xi_b^0\to\Sigma_b^+ K^-$&-17.0&$\Sigma_b^+\to\Xi_b^0 K^+$&-17.0&$\Xi_b^0\to\Xi_b^{\prime 0}\eta_8$&16.0&$\Xi_b^{\prime 0}\to\Xi_b^0\eta_8$&16.0\\
		\hline
		\rule{0pt}{12pt}$\Sigma_b^0\to\Lambda_b^0\pi^0$&18.5&$\Lambda_b^0\to\Sigma_b^+\pi^-$&-15.0&$\Sigma_b^+\to\Lambda_b^0\pi^+$&-15.0&$\Lambda_b^0\to\Sigma_b^-\pi^+$&15.0&$\Sigma_b^-\to\Lambda_b^0\pi^-$&15.0\\
		\hline
		\rule{0pt}{12pt}$\Xi_b^{\prime 0}\to\Lambda_b^0\bar{K^0}$&-11.5&$\Lambda_b^0\to\Xi_b^{\prime -}K^+$&11.5&$\Xi_b^{\prime -}\to\Lambda_b^0K^-$&11.5&$\Xi_b^-\to\Sigma_b^0K^-$&-11.5&$\Sigma_b^0\to\Xi_b^-K^+$&-11.5\\
		\hline
		\rule{0pt}{12pt}$\Sigma_b^-\to\Xi_b^-K^0$&-17.0&$\Xi_b^-\to\Xi_b^{\prime -}\eta_8$&16.0&$\Xi_b^{\prime -}\to\Xi_b^-\eta_8$&16.0&$\Xi_b^-\to\Omega_b^-K^0$&17.0&$\Omega_b^-\to\Xi_b^-\bar{K^0}$&17.0\\
		\hline
		\rule{0pt}{12pt}$\Xi_b^{\prime 0}\to\Xi_b^-\pi^+$&10.6&$\Xi_b^-\to\Xi_b^{\prime -}\pi^0$&-7.5&$\Xi_b^{\prime -}\to\Xi_b^-\pi^0$&-7.5&$\Xi_b^0\to\Omega_b^-K^+$&17.0&$\Omega_b^-\to\Xi_b^0K^-$&17.0\\
		\hline
		\rule{0pt}{12pt}$\Xi_b^{\prime 0}\to\Xi_b^0\pi^0$&7.5&$\Xi_b^0\to\Xi_b^{\prime -}\pi^+$&10.6&$\Xi_b^{\prime -}\to\Xi_b^0\pi^-$&10.6&$\Xi_b^{\prime 0}\to\Sigma_b^0\bar{K^0}$&13.4&$\Sigma_b^0\to\Xi_b^{\prime 0}K^0$&13.4\\
		\hline
		\rule{0pt}{12pt}$\Sigma_b^+\to\Xi_b^{\prime 0}K^+$&19.0&$\Xi_b^{\prime 0}\to\Xi_b^{\prime 0}\eta_8$&-5.3&$\Sigma_b^0\to\Sigma_b^0\eta_8$&12.5&$\Sigma_b^0\to\Sigma_b^+\pi^-$&17.0&$\Sigma_b^+\to\Sigma_b^0\pi^+$&17.0\\
		\hline
		\rule{0pt}{12pt}$\Sigma_b^+\to\Sigma_b^+\pi^0$&18.0&$\Xi_b^{\prime -}\to\Sigma_b^0K^-$&13.4&$\Sigma_b^0\to\Xi_b^{\prime -}K^+$&13.4&$\Xi_b^{\prime -}\to\Sigma_b^-\bar{K^0}$&19.0&$\Sigma_b^-\to\Xi_b^{\prime -}K^0$&19.0\\
		\hline
		\rule{0pt}{12pt}$\Omega_b^-\to\Omega_b^-\eta_8$&-26.0&$\Omega_b^-\to\Xi_b^{\prime 0}K^-$&21.0&$\Xi_b^{\prime 0}\to\Omega_b^-K^+$&21.0&$\Omega_b^-\to\Xi_b^{\prime -}\bar{K^0}$&19.0&$\Xi_b^{\prime-}\to\Omega_b^-K^0$&19.0\\
		\hline
		\rule{0pt}{12pt}$\Sigma_b^0\to\Sigma_b^-\pi^+$&17.0&$\Sigma_b^-\to\Sigma_b^-\eta_8$&12.5&$\Sigma_b^-\to\Sigma_b^-\pi^0$&-18.0&$\Xi_b^{\prime -}\to\Xi_b^{\prime 0}\pi^-$&12.0&$\Xi_b^{\prime 0}\to\Xi_b^{\prime -}\pi^+$&12.0\\
		\hline
		\rule{0pt}{12pt}$\Xi_b^{\prime 0}\to\Xi_b^{\prime 0}\pi^0$&9.0&$\Lambda_b^0\to\Xi_b^-K^+$&-1.5&$\Xi_b^0\to\Sigma_b^0\bar{K^0}$&-11.5&$\Lambda_b^0\to\Sigma_b^0\pi^0$&18.5&$\Lambda_b^0\to\Xi_b^{\prime 0}K^0$&-11.5\\
		\hline
		\rule{0pt}{12pt}$\Xi_b^-\to\Sigma_b^-\bar{K^0}$&-17.0&$\Xi_b^-\to\Xi_b^{\prime 0}\pi^-$&10.6&$\Xi_b^0\to\Xi_b^{\prime 0}\pi^0$&7.5&$\Xi_b^{\prime 0}\to\Sigma_b^+K^-$&19.0&$\Sigma_b^+\to\Sigma_b^+\eta_8$&12.5\\
		\hline
		\rule{0pt}{12pt}$\Xi_b^{\prime -}\to\Xi_b^{\prime -}\eta_8$&-5.5&$\Sigma_b^-\to\Sigma_b^0\pi^-$&17.0&$\Xi_b^{\prime -}\to\Xi_b^{\prime -}\pi^0$&-9.0&&\\
		\hline
		\hline
	\end{tabular}
\end{table}

\begin{table}[h]
	\centering
	\caption{Strong coupling constants of $V\mathcal{B}_{\bar{3}}\mathcal{B}_{\bar{3}}$, $V\mathcal{B}_{\bar{3}}\mathcal{B}_6$ and $V\mathcal{B}_6\mathcal{B}_6$ vertex.}
	\label{ap:VBB}
	\begin{tabular}{|c|c|c|c|c|c|c|c|c|c|c|c|}
		\hline
		\hline
		\rule{0pt}{12pt}Vertex&$f_1$&$f_2$&Vertex&$f_1$&$f_2$&Vertex&$f_1$&$f_2$&Vertex&$f_1$&$f_2$\\
		\hline
		\hline
		\rule{0pt}{12pt}$\Lambda_b^0\to\Lambda_b^0\omega$&5.2&8.0&$\Lambda_b^0\to\Xi_b^0K^{\ast 0}$&5.0&7.0&$\Xi_b^0\to\Lambda_b^0\bar{K^{\ast 0}}$&5.0&7.0&$\Xi_b^0\to\Xi_b^0\phi$&4.0&5.7\\
		\hline
		\rule{0pt}{12pt}$\Xi_b^-\to\Lambda_b^0K^{\ast -}$&-5.0&-7.0&$\Xi_b^-\to\Xi_b^-\phi$&4.0&5.7&$\Xi_b^-\to\Xi_b^0\rho^-$&4.4&7.1&$\Xi_b^0\to\Xi_b^-\rho^+$&4.4&7.1\\
		\hline
		\rule{0pt}{12pt}$\Xi_b^-\to\Xi_b^-\rho^0$&-3.1&-5.0&$\Xi_b^0\to\Xi_b^0\omega$&2.8&4.0&$\Xi_b^0\to\Xi_b^0\rho^0$&3.1&5.0&$\Lambda_b^0\to\Sigma_b^0\rho^0$&2.8&40.0\\
		\hline
		\rule{0pt}{12pt}$\Lambda_b^0\to\Sigma_b^+\rho^-$&-2.8&-40.0&$\Sigma_b^+\to\Lambda_b^0\rho^+$&-2.8&-40.0&$\Lambda_b^0\to\Xi_b^{\prime 0}K^{\ast 0}$&-2.6&-32.0&$\Xi_b^{\prime 0}\to\Lambda_b^0\bar{K^{\ast 0}}$&-2.6&-32.0\\
		\hline
		\rule{0pt}{12pt}$\Sigma_b^0\to\Xi_b^0K^{\ast 0}$&-2.6&-32.0&$\Xi_b^0\to\Sigma_b^+K^{\ast -}$&-3.7&-45.3&$\Sigma_b^+\to\Xi_b^0K^{\ast +}$&-3.7&-45.3&$\Xi_b^0\to\Xi_b^{\prime 0}\phi$&-2.6&-30.0\\
		\hline
		\rule{0pt}{12pt}$\Lambda_b^0\to\Sigma_b^-\rho^+$&2.8&40.0&$\Sigma_b^-\to\Lambda_b^0\rho^-$&2.8&40.0&$\Lambda_b^0\to\Xi_b^{\prime -}K^{\ast +}$&2.6&32.0&$\Xi_b^{\prime -}\to\Lambda_b^0K^{\ast -}$&2.6&32.0\\
		\hline
		\rule{0pt}{12pt}$\Sigma_b^0\to\Xi_b^-K^{\ast +}$&-2.6&-32.0&$\Xi_b^-\to\Sigma_b^-\bar{K^{\ast 0}}$&-3.7&-45.3&$\Sigma_b^-\to\Xi_b^-K^{\ast 0}$&-3.7&-45.3&$\Xi_b^-\to\Xi_b^{\prime -}\phi$&-2.6&-30.0\\
		\hline
		\rule{0pt}{12pt}$\Xi_b^-\to\Omega_b^-K^{\ast 0}$&3.5&50.0&$\Omega_b^-\to\Xi_b^-\bar{K^{\ast 0}}$&3.5&50.0&$\Xi_b^-\to\Xi_b^{\prime 0}\rho^-$&2.0&28.3&$\Xi_b^{\prime 0}\to\Xi_b^-\rho^+$&2.0&28.3\\
		\hline
		\rule{0pt}{12pt}$\Xi_b^{\prime -}\to\Xi_b^-\omega$&1.4&22.0&$\Xi_b^-\to\Xi_b^{\prime -}\rho^0$&-1.4&-20.0&$\Xi_b^{\prime -}\to\Xi_b^-\rho^0$&-1.4&-20.0&$\Xi_b^0\to\Omega_b^-K^{\ast +}$&3.7&45.3\\
		\hline
		\rule{0pt}{12pt}$\Xi_b^0\to\Xi_b^{\prime -}\rho^+$&2.0&28.3&$\Xi_b^{\prime-}\to\Xi_b^0\rho^-$&2.0&28.3&$\Xi_b^0\to\Xi_b^{\prime 0}\omega$&1.4&22.0&$\Xi_b^{\prime 0}\to\Xi_b^0\omega$&1.4&22.0\\
		\hline
		\rule{0pt}{12pt}$\Xi_b^{\prime 0}\to\Xi_b^0\rho^0$&1.4&22.0&$\Sigma_b^0\to\Sigma_b^0\omega$&4.0&50.0&$\Sigma_b^0\to\Sigma_b^{\prime +}\rho^-$&4.5&55.0&$\Sigma_b^+\to\Sigma_b^0\rho^+$&4.5&55.0\\
		\hline
		\rule{0pt}{12pt}$\Xi_b^{\prime 0}\to\Sigma_b^0\bar{K^{\ast 0}}$&4.2&42.4&$\Sigma_b^+\to\Sigma_b^+\omega$&4.0&50.0&$\Sigma_b^+\to\Sigma_b^+\rho^0$&4.4&60.0&$\Sigma_b^+\to\Xi_b^{\prime 0}K^{\ast +}$&6.0&60.0\\
		\hline
		\rule{0pt}{12pt}$\Xi_b^{\prime 0}\to\Xi_b^{\prime 0}\phi$&5.0&45.0&$\Omega_b^-\to\Omega_b^-\phi$&10.0&95.0&$\Omega_b^-\to\Xi_b^{\prime 0}K^{\ast -}$&6.0&70.0&$\Xi_b^{\prime 0}\to\Omega_b^-K^{\ast +}$&6.0&70.0\\
		\hline
		\rule{0pt}{12pt}$\Xi_b^{\prime -}\to\Omega_b^-K^{\ast 0}$&6.0&60.0&$\Sigma_b^-\to\Sigma_b^0\rho^-$&4.5&55.0&$\Sigma_b^0\to\Sigma_b^-\rho^+$&4.5&55.0&$\Sigma_b^-\to\Sigma_b^-\omega$&4.0&50.0\\
		\hline
		\rule{0pt}{12pt}$\Sigma_b^-\to\Xi_b^{\prime -}K^{\ast 0}$&6.0&60.0&$\Xi_b^{\prime -}\to\Sigma_b^-\bar{K^{\ast 0}}$&6.0&60.0&$\Sigma_b^0\to\Xi_b^{\prime -}K^{\ast +}$&4.2&42.4&$\Xi_b^{\prime -}\to\Sigma_b^0K^{\ast -}$&4.2&42.4\\
		\hline
		\rule{0pt}{12pt}$\Xi_b^{\prime -}\Xi_b^{\prime 0}\rho^-$&3.2&38.9&$\Xi_b^{\prime 0}\to\Xi_b^{\prime -}\rho^+$&3.2&38.9&$\Xi_b^{\prime -}\to\Xi_b^{\prime -}\omega$&2.0&25.0&$\Xi_b^{\prime -}\to\Xi_b^{\prime -}\rho^0$&-2.2&-30.0\\
		\hline
		\rule{0pt}{12pt}$\Xi_b^{\prime 0}\to\Xi_b^{\prime 0}\rho^0$&2.2&30.0&$\Lambda_b^0\to\Xi_b^-K^{\ast +}$&-5.0&-7.0&$\Xi_b^-\to\Xi_b^-\omega$&2.8&4.0&$\Sigma_b^0\to\Lambda_b^0\rho^0$&2.8&40.0\\
		\hline
		\rule{0pt}{12pt}$\Xi_b^0\to\Sigma_b^0\bar{K^{\ast 0}}$&-2.6&-32.0&$\Xi_b^{\prime 0}\to\Xi_b^0\phi$&-2.6&-30.0&$\Xi_b^-\to\Sigma_b^0K^{\ast -}$&-2.6&-32.0&$\Xi_b^{\prime -}\to\Xi_b^-\phi$&-2.6&-30.0\\
		\hline
		\rule{0pt}{12pt}$\Xi_b^-\to\Xi_b^{\prime -}\omega$&1.4&22.0&$\Omega_b^-\to\Xi_b^0K^{\ast -}$&3.7&45.3&$\Xi_b^0\to\Xi_b^{\prime 0}\rho^0$&1.4&22.0&$\Sigma_b^0\to\Xi_b^{\prime 0}K^{\ast 0}$&4.2&42.4\\
		\hline
		\rule{0pt}{12pt}$\Xi_b^{\prime 0}\to\Sigma_b^+K^{\ast -}$&6.0&60.0&$\Omega_b^-\to\Xi_b^{\prime -}\bar{K^{\ast 0}}$&6.0&60.0&$\Sigma_b^-\to\Sigma_b^-\rho^0$&-4.4&-60.0&$\Xi_b^{\prime -}\to\Xi_b^{\prime -}\phi$&5.0&45.0\\
		\hline
		\rule{0pt}{12pt}$\Xi_b^{\prime 0}\to\Xi_b^{\prime 0}\omega$&2.1&27.0&&&&&\\
		\hline
		\hline
	\end{tabular}
\end{table}


\section{ Expressions of amplitudes}\label{app:amp}

The expressions of amplitudes for all the $\mathcal{B}_{bc}\to\mathcal{B}_bP$ decays are collected in this Appendix.
{\scriptsize
	\begin{align}
	\mathcal{A}(\Xi_{bc}^+\to\Lambda_b^0K^+)=&\mathcal{T}(\Xi_{bc}^+\to\Lambda_b^0K^+)+\mathcal{M}(K^+,\Lambda_b^0;\omega)+\mathcal{M}(K^+,\Sigma_b^0;\rho^0)+\mathcal{M}(K^{\ast+},\Lambda_b^0;\eta_8)+\mathcal{M}(K^{\ast +},\Sigma_b^0;\pi^0)\non
	&+\mathcal{M}(K^+,\Lambda_b^0;\Xi_b^-)+\mathcal{M}(K^+,\Lambda_b^0;\Xi_b^{\prime -})+\mathcal{M}(K^+,\Sigma_b^0;\Xi_b^-)+\mathcal{M}(K^+,\Sigma_b^0;\Xi_b^{\prime -})+\mathcal{M}(K^{\ast +},\Lambda_b^0;\Xi_b^-)\non
	&+\mathcal{M}(K^{\ast +},\Lambda_b^0;\Xi_b^{\prime -})+\mathcal{M}(K^{\ast +},\Sigma_b^0;\Xi_b^-)
	+\mathcal{M}(K^{\ast +},\Sigma_b^0;\Xi_b^{\prime -}),\\
	\mathcal{A}(\Xi_{bc}^+\to\Lambda_b^0\pi^+)=&\mathcal{T}(\Xi_{bc}^+\to\Lambda_b^0\pi^+)+\mathcal{M}(\pi^+,\Sigma_b^0;\rho^0)+\mathcal{M}(\rho^+,\Sigma_b^0;\pi^0)+\mathcal{M}(K^+,\Xi_b^0;K^{\ast 0})+\mathcal{M}(K^+,\Xi_b^{\prime 0};K^{\ast 0})\non
	&+\mathcal{M}(K^{\ast +},\Xi_b^0;K^0)+\mathcal{M}(K^{\ast +},\Xi_b^{\prime 0};K^0)+\mathcal{M}(\pi^+,\Lambda_b^0;\Sigma_b^-)+\mathcal{M}(\pi^+,\Sigma_b^0;\Sigma_b^-)+\mathcal{M}(\rho^+,\Lambda_b^0;\Sigma_b^-)\non
	&+\mathcal{M}(\rho^+,\Sigma_b^0;\Sigma_b^-)+\mathcal{M}(K^+,\Xi_b^0;\Xi_b^-)+\mathcal{M}(K^+,\Xi_b^0;\Xi_b^{\prime -})+\mathcal{M}(K^+,\Xi_b^{\prime 0};\Xi_b^-)+\mathcal{M}(K^+,\Xi_b^{\prime 0};\Xi_b^{\prime -})\non
	&+\mathcal{M}(K^{\ast +},\Xi_b^0;\Xi_b^-)+\mathcal{M}(K^{\ast +},\Xi_b^0;\Xi_b^{\prime -})+\mathcal{M}(K^{\ast +},\Xi_b^{\prime 0};\Xi_b^-)
	+\mathcal{M}(K^{\ast +},\Xi_b^{\prime 0};\Xi_b^{\prime -}),\\
	\mathcal{A}(\Xi_{bc}^+\to\Xi_b^0K^+)=&\mathcal{T}(\Xi_{bc}^+\to\Xi_b^0K^+)+\mathcal{M}(K^+,\Xi_b^0;\rho^0)+\mathcal{M}(K^+,\Xi_b^0;\omega)+\mathcal{M}(K^+,\Xi_b^{\prime 0};\rho^0)+\mathcal{M}(K^+,\Xi_b^{\prime 0};\omega)\non
	&+\mathcal{M}(K^{\ast +},\Xi_b^0;\pi^0)+\mathcal{M}(K^{\ast +},\Xi_b^0;\eta_8)+\mathcal{M}(K^{\ast +},\Xi_b^{\prime 0};\pi^0)+\mathcal{M}(K^{\ast +},\Xi_b^{\prime 0};\eta_8)+\mathcal{M}(\pi^+,\Lambda_b^0;\bar{K}^{\ast 0})\non
	&+\mathcal{M}(\pi^+,\Sigma_b^0;\bar{K}^{\ast 0})+\mathcal{M}(\rho^+,\Lambda_b^0;\bar{K}^{0})+\mathcal{M}(\rho^+,\Sigma_b^0;\bar{K}^{0})+\mathcal{M}(K^+,\Xi_b^0;\phi)
	+\mathcal{M}(K^+,\Xi_b^{\prime 0};\phi)\non
	&+\mathcal{M}(K^{\ast +},\Xi_b^0;\eta_8)+\mathcal{M}(K^{\ast +},\Xi_b^{\prime 0};\eta_8)+\mathcal{M}(K^{+},\Xi_b^0;\Omega_b^-)+\mathcal{M}(K^{+},\Xi_b^{\prime 0};\Omega_b^-)
	+\mathcal{M}(K^{\ast +},\Xi_b^0;\Omega_b^-)\non
	&+\mathcal{M}(K^{\ast +},\Xi_b^{\prime 0};\Omega_b^-)+\mathcal{M}(\pi^{+},\Lambda_b^0;\Xi_b^-)+\mathcal{M}(\pi^{+},\Lambda_b^0;\Xi_b^{\prime -})+\mathcal{M}(\pi^{+},\Sigma_b^0;\Xi_b^-)
	+\mathcal{M}(\pi^{+},\Sigma_b^0;\Xi_b^{\prime -})\non
	&+\mathcal{M}(\rho^{+},\Lambda_b^0;\Xi_b^-)+\mathcal{M}(\rho^{+},\Lambda_b^0;\Xi_b^{\prime -})+\mathcal{M}(\rho^{+},\Sigma_b^0;\Xi_b^-)
	+\mathcal{M}(\rho^{+},\Sigma_b^0;\Xi_b^{\prime -}),\\
	\mathcal{A}(\Xi_{bc}^+\to\Xi_b^0\pi^+)=&\mathcal{T}(\Xi_{bc}^+\to\Xi_b^0\pi^+)+\mathcal{M}(\pi^+,\Xi_b^0;\rho^0)+\mathcal{M}(\pi^+,\Xi_b^{\prime 0};\rho^0)+\mathcal{M}(\rho^+,\Xi_b^0;\pi^0)
	+\mathcal{M}(\rho^+,\Xi_b^{\prime 0};\pi^0)\non
	&+\mathcal{M}(\pi^+,\Xi_b^0;\Xi_b^-)+\mathcal{M}(\pi^+,\Xi_b^0;\Xi_b^{\prime -})+\mathcal{M}(\pi^+,\Xi_b^{\prime 0};\Xi_b^-)+\mathcal{M}(\pi^+,\Xi_b^{\prime 0};\Xi_b^{\prime -})
	+\mathcal{M}(\rho^+,\Xi_b^0;\Xi_b^-)\non
	&+\mathcal{M}(\rho^+,\Xi_b^0;\Xi_b^{\prime -})+\mathcal{M}(\rho^+,\Xi_b^{\prime 0};\Xi_b^-)
	+\mathcal{M}(\rho^+,\Xi_b^{\prime 0};\Xi_b^{\prime -}),  \\
	\mathcal{A}(\Xi_{bc}^0\to\Lambda_b^0\eta_8)=&\mathcal{C}_{SD}(\Xi_{bc}^0\to\Lambda_b^0\eta_8)+\mathcal{M}(K^+,\Xi_b^-;K^{\ast +})+\mathcal{M}(K^+,\Xi_b^{\prime -};K^{\ast +})+\mathcal{M}(K^{\ast +},\Xi_b^-;K^+)
	+\mathcal{M}(K^{\ast +},\Xi_b^{\prime -};K^+)\non
	&+\mathcal{M}(K^+,\Xi_b^-;\Xi_b^-)+\mathcal{M}(K^+,\Xi_b^-;\Xi_b^{\prime -})+\mathcal{M}(K^+,\Xi_b^{\prime -};\Xi_b^-)+\mathcal{M}(K^+,\Xi_b^{\prime -};\Xi_b^{\prime -})
	+\mathcal{M}(K^{\ast +},\Xi_b^-;\Xi_b^-)\non
	&+\mathcal{M}(K^{\ast +},\Xi_b^-;\Xi_b^{\prime -})+\mathcal{M}(K^{\ast +},\Xi_b^{\prime -};\Xi_b^-)+\mathcal{M}(K^{\ast +},\Xi_b^{\prime -};\Xi_b^{\prime -})+\mathcal{M}(\pi^+,\Sigma_b^-;\Sigma_b^-)
	+\mathcal{M}(\rho^+,\Sigma_b^-;\Sigma_b^-),\\
	\mathcal{A}(\Xi_{bc}^0\to\Lambda_b^0\bar{K}^0		)=&\mathcal{C}_{SD}(\Xi_{bc}^0\to\Lambda_b^0\bar{K}^0)+\mathcal{M}(\pi^+,\Xi_b^-;K^{\ast +})+\mathcal{M}(\pi^+,\Xi_b^{\prime -};K^{\ast +})+\mathcal{M}(\rho^+,\Xi_b^-;K^+)+\mathcal{M}(\rho^+,\Xi_b^{\prime -};K^+)\non
	&+\mathcal{M}(\pi^+,\Xi_b^-;\Sigma_b^-)+\mathcal{M}(\pi^+,\Xi_b^{\prime -};\Sigma_b^-)+\mathcal{M}(\rho^+,\Xi_b^-;\Sigma_b^-)
	+\mathcal{M}(\rho^+,\Xi_b^{\prime -};\Sigma_b^-),\\
	\mathcal{A}(\Xi_{bc}^0\to\Lambda_b^0K^0)=&\mathcal{C}_{SD}(\Xi_{bc}^0\to\Lambda_b^0K^0)+\mathcal{M}(K^+,\Sigma_b^-;\rho^+)+\mathcal{M}(K^{\ast +},\Sigma_b^-;\pi^+)+\mathcal{M}(K+,\Sigma_b^-;\Xi_b^-)+\mathcal{M}(K^+,\Sigma_b^-;\Xi_b^{\prime -})\non
	&+\mathcal{M}(K^{\ast +},\Sigma_b^-;\Xi_b^-)+\mathcal{M}(K^{\ast +},\Sigma_b^-;\Xi_b^{\prime -}),\\
	\mathcal{A}(\Xi_{bc}^0\to\Lambda_b^0\pi^0)=&\mathcal{C}_{SD}(\Xi_{bc}^0\to\Lambda_b^0\pi^0)+\mathcal{M}(\pi^+,\Sigma_b^-;\rho^+)+\mathcal{M}(\rho^+,\Sigma_b^-;\pi^+)+\mathcal{M}(K^{+},\Xi_b^-;K^{\ast +})
	+\mathcal{M}(K^{+},\Xi_b^{\prime -};K^{\ast +})\non
	&+\mathcal{M}(K^{\ast +},\Xi_b^-;K^{+})+\mathcal{M}(K^{\ast +},\Xi_b^{\prime -};K^{+})+\mathcal{M}(\pi^+,\Sigma_b^-;\Sigma_b^-)+\mathcal{M}(\rho^+,\Sigma_b^-;\Sigma_b^-)+\mathcal{M}(K^+,\Xi_b^-;\Xi_b^-)\non
	&+\mathcal{M}(K^+,\Xi_b^-;\Xi_b^{\prime -})+\mathcal{M}(K^+,\Xi_b^{\prime -};\Xi_b^-)+\mathcal{M}(K^+,\Xi_b^{\prime -};\Xi_b^{\prime -})\\
	\mathcal{A}(\Xi_{bc}^0\to\Xi_b^-K^+)=&\mathcal{T}(\Xi_{bc}^0\to\Xi_b^-K^+)+\mathcal{M}(\pi^+,\Sigma_b^-;\bar{K}^{\ast 0})+\mathcal{M}(\rho^+,\Sigma_b^-;\bar{K}^{0})+\mathcal{M}(K^+,\Xi_b^-;\phi)
	+\mathcal{M}(K^+,\Xi_b^{\prime -};\phi)\non
	&+\mathcal{M}(K^{\ast +},\Xi_b^-;\eta_8)+\mathcal{M}(K^{\ast +},\Xi_b^{\prime -};\eta_8),\\
	\mathcal{A}(\Xi_{bc}^0\to\Xi_b^-\pi^+)=&\mathcal{T}(\Xi_{bc}^0\to\Xi_b^-\pi^+)+\mathcal{M}(\pi^+,\Xi_b^-;\rho^0)+\mathcal{M}(\pi^+,\Xi_b^{\prime -};\rho^0)
	+\mathcal{M}(\rho^+,\Xi_b^-;\pi^0)
	+\mathcal{M}(\rho^+,\Xi_b^{\prime -};\pi^0)\\
	\mathcal{A}(\Xi_{bc}^0\to\Xi_b^0\eta_8)=&\mathcal{M}(\pi^+,\Xi_b^-;\Xi_b^-)+\mathcal{M}(\pi^+,\Xi_b^-;\Xi_b^{\prime -})+\mathcal{M}(\pi^+,\Xi_b^{\prime -};\Xi_b^-)+\mathcal{M}(\pi^+,\Xi_b^{\prime -};\Xi_b^{\prime -})
	+\mathcal{M}(\rho^+,\Xi_b^-;\Xi_b^-)\non
	&+\mathcal{M}(\rho^+,\Xi_b^-;\Xi_b^{\prime -})+\mathcal{M}(\rho^+,\Xi_b^{\prime -};\Xi_b^-)+\mathcal{M}(\rho^+,\Xi_b^{\prime -};\Xi_b^{\prime -})\\
	\mathcal{A}(\Xi_{bc}^0\to\Xi_b^0K^0)=&\mathcal{C}_{SD}(\Xi_{bc}^0\to\Xi_b^0K^0)+\mathcal{M}(K^+,\Xi_b^-;\rho^+)+\mathcal{M}(K^+,\Xi_b^{\prime -};\rho^+)+\mathcal{M}(K^{\ast +},\Xi_b^-;\pi^+)
	+\mathcal{M}(K^{\ast +},\Xi_b^{\prime -};\pi^+)\non
	&+\mathcal{M}(K^{+},\Xi_b^-;\Omega_b^-)+\mathcal{M}(K^{+},\Xi_b^{\prime -};\Omega_b^-)+\mathcal{M}(K^{\ast +},\Xi_b^-;\Omega_b^-)+\mathcal{M}(K^{\ast +},\Xi_b^{\prime -};\Omega_b^-)
	+\mathcal{M}(\pi^{+},\Sigma_b^-;\Xi_b^-)\non
	&+\mathcal{M}(\pi^{+},\Sigma_b^-;\Xi_b^{\prime -})+\mathcal{M}(\rho^{+},\Sigma_b^-;\Xi_b^-)+\mathcal{M}(\rho^{+},\Sigma_b^-;\Xi_b^{\prime -}),\\
	\mathcal{A}(\Xi_{bc}^0\to\Xi_b^0\pi^0)=&\mathcal{C}_{SD}(\Xi_{bc}^0\to\Xi_b^0\pi^0)+\mathcal{M}(\pi^{+},\Xi_b^-;\rho^+)+\mathcal{M}(\pi^{+},\Xi_b^{\prime -};\rho^+)+\mathcal{M}(\rho^{+},\Xi_b^-;\pi^+)
	+\mathcal{M}(\rho^{+},\Xi_b^{\prime -};\pi^+)\non
	&+\mathcal{M}(\pi^{+},\Xi_b^-;\Xi_b^-)+\mathcal{M}(\pi^{+},\Xi_b^-;\Xi_b^{\prime -})+\mathcal{M}(\pi^{+},\Xi_b^{\prime -};\Xi_b^-)+\mathcal{M}(\pi^{+},\Xi_b^{\prime -};\Xi_b^{\prime -})
	+\mathcal{M}(\rho^{+},\Xi_b^-;\Xi_b^-)\non
	&+\mathcal{M}(\rho^{+},\Xi_b^-;\Xi_b^{\prime -})+\mathcal{M}(\rho^{+},\Xi_b^{\prime -};\Xi_b^-)+\mathcal{M}(\rho^{+},\Xi_b^{\prime -};\Xi_b^{\prime -})\\
	\mathcal{A}(\Xi_{bc}^0\to\Sigma_b^-K^+)=&\mathcal{T}(\Xi_{bc}^0\to\Sigma_b^-K^+),\\
	\mathcal{A}(\Omega_{bc}^0\to\Omega_b^-\pi^+)=&\mathcal{T}(\Omega_{bc}^0\to\Omega_b^-\pi^+),\\
	\mathcal{A}(\Omega_{bc}^0\to\Lambda_b^0\eta_8)=&\mathcal{C}_{SD}(\Omega_{bc}^0\to\Lambda_b^0\eta_8)+\mathcal{M}(K^{+},\Xi_b^-;K^{\ast +})+\mathcal{M}(K^{+},\Xi_b^{\prime -};K^{\ast +})+\mathcal{M}(K^{\ast +},\Xi_b^-;K^{+})
	+\mathcal{M}(K^{\ast +},\Xi_b^{\prime -};K^{+})\non
	&+\mathcal{M}(K^{+},\Xi_b^-;\Xi_b^-)+\mathcal{M}(K^{+},\Xi_b^-;\Xi_b^{\prime -})+\mathcal{M}(K^{+},\Xi_b^{\prime -};\Xi_b^-)+\mathcal{M}(K^{+},\Xi_b^{\prime -};\Xi_b^{\prime -})
	+\mathcal{M}(K^{\ast +},\Xi_b^-;\Xi_b^-)\non
	&+\mathcal{M}(K^{\ast +},\Xi_b^-;\Xi_b^{\prime -})+\mathcal{M}(K^{\ast +},\Xi_b^{\prime -};\Xi_b^-)
	+\mathcal{M}(K^{\ast +},\Xi_b^{\prime -};\Xi_b^{\prime -}, )\\
	\mathcal{A}(\Omega_{bc}^0\to\Lambda_b^0\bar{K}^0)=&\mathcal{C}_{SD}(\Omega_{bc}^0\to\Lambda_b^0\bar{K}^0)+\mathcal{M}(\pi^+,\Xi_b^-;K^{\ast +})+\mathcal{M}(\pi^+,\Xi_b^{\prime -};K^{\ast +})+\mathcal{M}(\rho^+,\Xi_b^-;K^{+})
	+\mathcal{M}(\rho^+,\Xi_b^{\prime -};K^{+})\non
	&+\mathcal{M}(\pi^+,\Xi_b^-;\Sigma_b^-)+\mathcal{M}(\pi^+,\Xi_b^{\prime -};\Sigma_b^-)+\mathcal{M}(\rho^+,\Xi_b^-;\Sigma_b^-)+\mathcal{M}(\rho^+,\Xi_b^{\prime -};\Sigma_b^-)
	+\mathcal{M}(K^+,\Omega_b^-;\Xi_b^-)\non
	&+\mathcal{M}(K^+,\Omega_b^-;\Xi_b^{\prime -})+\mathcal{M}(K^{\ast +},\Omega_b^-;\Xi_b^-)
	+\mathcal{M}(K^{\ast +},\Omega_b^-;\Xi_b^{\prime -}),\\
	\mathcal{A}(\Omega_{bc}^0\to\Lambda_b^0\pi^0)=&\mathcal{M}(K^{+},\Xi_b^-;K^{\ast +})+\mathcal{M}(K^{+},\Xi_b^{\prime -};K^{\ast +})+\mathcal{M}(K^{\ast +},\Xi_b^-;K^{+})
	+\mathcal{M}(K^{\ast +},\Xi_b^{\prime -};K^{+})\non
	&+\mathcal{M}(K^{+},\Xi_b^-;\Xi_b^-)+\mathcal{M}(K^{+},\Xi_b^-;\Xi_b^{\prime -})+\mathcal{M}(K^{+},\Xi_b^{\prime -};\Xi_b^-)
	+\mathcal{M}(K^{+},\Xi_b^{\prime -};\Xi_b^{\prime -})\non
	&+\mathcal{M}(K^{\ast +},\Xi_b^-;\Xi_b^-)+\mathcal{M}(K^{\ast +},\Xi_b^-;\Xi_b^{\prime -})+\mathcal{M}(K^{\ast +},\Xi_b^{\prime -};\Xi_b^-)
	+\mathcal{M}(K^{\ast +},\Xi_b^{\prime -};\Xi_b^{\prime -}),\\
	\mathcal{A}(\Omega_{bc}^0\to\Xi_b^-K^+)=&\mathcal{T}(\Omega_{bc}^0\to\Xi_b^-K^+)+\mathcal{M}(K^{+},\Xi_b^-;\phi)+\mathcal{M}(K^{+},\Xi_b^{\prime -};\phi)+\mathcal{M}(K^{\ast +},\Xi_b^-;\eta_8)
	+\mathcal{M}(K^{\ast +},\Xi_b^{\prime -};\eta_8),\\
	\mathcal{A}(\Omega_{bc}^0\to\Xi_b^-\pi^+)=&\mathcal{T}(\Omega_{bc}^0\to\Xi_b^-\pi^+)+\mathcal{M}(\pi^+,\Xi_b^-;\rho^0)+\mathcal{M}(\pi^+,\Xi_b^{\prime -};\rho^0)+\mathcal{M}(\rho^+,\Xi_b^-;\pi^0)
	+\mathcal{M}(\rho^+,\Xi_b^{\prime -};\pi^0)\non
	&+\mathcal{M}(K^+,\Omega_b^-;K^{\ast 0})+\mathcal{M}(K^{\ast +},\Omega_b^-;K^{0}), \\
	\mathcal{A}(\Omega_{bc}^0\to\Xi_b^0\eta_8)=&\mathcal{C}_{SD}(\Omega_{bc}^0\to\Xi_b^0\eta_8)+\mathcal{M}(K^+,\Omega_b^-;K^{\ast +})+\mathcal{M}(K^{\ast +},\Omega_b^-;K^{+})+\mathcal{M}(K^+,\Omega_b^-;\Omega_b^-)
	+\mathcal{M}(K^{\ast +},\Omega_b^-;\Omega_b^-)\non
	&+\mathcal{M}(\pi^+,\Xi_b^-;\Xi_b^-)+\mathcal{M}(\pi^+,\Xi_b^-;\Xi_b^{\prime -})+\mathcal{M}(\pi^+,\Xi_b^{\prime -};\Xi_b^-)+\mathcal{M}(\pi^+,\Xi_b^{\prime -};\Xi_b^{\prime -})
	+\mathcal{M}(\rho^+,\Xi_b^-;\Xi_b^-)\non
	&+\mathcal{M}(\rho^+,\Xi_b^-;\Xi_b^{\prime -})+\mathcal{M}(\rho^+,\Xi_b^{\prime -};\Xi_b^-)+\mathcal{M}(\rho^+,\Xi_b^{\prime -};\Xi_b^{\prime -}), \\
	\mathcal{A}(\Omega_{bc}^0\to\Xi_b^0\bar{K}^0)=&\mathcal{C}_{SD}(\Omega_{bc}^0\to\Xi_b^0\bar{K}^0)+\mathcal{M}(\pi^+,\Omega_b^-;K^{\ast +})+\mathcal{M}(\rho^+,\Omega_b^-;K^+)+\mathcal{M}(\pi^+,\Omega_b^-;\Xi_b^-)
	+\mathcal{M}(\pi^+,\Omega_b^-;\Xi_b^{\prime -})\non
	&+\mathcal{M}(\rho^+,\Omega_b^-;\Xi_b^-)+\mathcal{M}(\rho^+,\Omega_b^-;\Xi_b^{\prime -}),\\
	\mathcal{A}(\Omega_{bc}^0\to\Xi_b^0K^0)=&\mathcal{C}_{SD}(\Omega_{bc}^0\to\Xi_b^0K^0)+\mathcal{M}(K^+,\Xi_b^-;\rho^+)+\mathcal{M}(K^+,\Xi_b^{\prime -};\rho^+)+\mathcal{M}(K^{\ast +},\Xi_b^-;\pi^+)+\mathcal{M}(K^{\ast +},\Xi_b^{\prime -};\pi^+)\non
	&+\mathcal{M}(K^{+},\Xi_b^-;\Omega_b^-)+\mathcal{M}(K^{+},\Xi_b^-{\prime -};\Omega_b^-)+\mathcal{M}(K^{\ast +},\Xi_b^-;\Omega_b^-)+\mathcal{M}(K^{\ast +},\Xi_b^{\prime -};\Omega_b^-)\\
	\mathcal{A}(\Omega_{bc}^0\to\Xi_b^0\pi^0)=&\mathcal{C}_{SD}(\Omega_{bc}^0\to\Xi_b^0\pi^0)+\mathcal{M}(\pi^{+},\Xi_b^-;\rho^+)+\mathcal{M}(\pi^{+},\Xi_b^{\prime -};\rho^+)+\mathcal{M}(\rho^{+},\Xi_b^-;\pi^+)
	+\mathcal{M}(\rho^{+},\Xi_b^{\prime -};\pi^+)\non
	&+\mathcal{M}(K^{+},\Omega_b^-;K^{\ast +})+\mathcal{M}(K^{\ast +},\Omega_b^-;K^{+})+\mathcal{M}(\pi^+,\Xi_b^-;\Xi_b^-)+\mathcal{M}(\pi^+,\Xi_b^-;\Xi_b^{\prime -})
	+\mathcal{M}(\pi^+,\Xi_b^{\prime -};\Xi_b^-)\non
	&+\mathcal{M}(\pi^+,\Xi_b^{\prime -};\Xi_b^{\prime -})+\mathcal{M}(\rho^+,\Xi_b^-;\Xi_b^-)+\mathcal{M}(\rho^+,\Xi_b^-;\Xi_b^{\prime -})+\mathcal{M}(\rho^+,\Xi_b^{\prime -};\Xi_b^-)
	+\mathcal{M}(\rho^+,\Xi_b^{\prime -};\Xi_b^{\prime -}), \\
	\mathcal{A}(\Xi_{bc}^+\to\Sigma_b^0K^+)=&\mathcal{T}(\Xi_{bc}^+\to\Sigma_b^0K^+)+\mathcal{M}(K^{+},\Lambda_b^0;\rho^0)+\mathcal{M}(K^{+},\Sigma_b^0;\omega)+\mathcal{M}(K^{\ast +},\Lambda_b^0;\pi^0)
	+\mathcal{M}(K^{\ast +},\Sigma_b^0;\eta_8)\non
	&+\mathcal{M}(K^{+},\Lambda_b^0;\Xi_b^-)+\mathcal{M}(K^{+},\Lambda_b^0;\Xi_b^{\prime -})+\mathcal{M}(K^{+},\Sigma_b^0;\Xi_b^-)+\mathcal{M}(K^{+},\Sigma_b^0;\Xi_b^{\prime -})
	+\mathcal{M}(K^{\ast +},\Lambda_b^0;\Xi_b^-)\non
	&+\mathcal{M}(K^{\ast +},\Lambda_b^0;\Xi_b^{\prime})+\mathcal{M}(K^{\ast +},\Sigma_b^0;\Xi_b^-)+\mathcal{M}(K^{\ast +},\Sigma_b^0;\Xi_b^{\prime -}),\\
	\mathcal{A}(\Xi_{bc}^+\to\Sigma_b^0\pi^+)=&\mathcal{T}(\Xi_{bc}^+\to\Sigma_b^0\pi^+)+\mathcal{M}(\pi^{+},\Lambda_b^0;\rho^0)+\mathcal{M}(\rho^{+},\Lambda_b^0;\pi^0)+\mathcal{M}(K^{+},\Xi_b^0;K^{\ast 0})
	+\mathcal{M}(K^{+},\Xi_b^{\prime 0};K^{\ast 0})\non
	&+\mathcal{M}(K^{\ast +},\Xi_b^0;K^{0})+\mathcal{M}(K^{\ast +},\Xi_b^{\prime 0};K^{0})+\mathcal{M}(\pi^+,\Lambda_b^0;\Sigma_b^-)+\mathcal{M}(\pi^+,\Sigma_b^0;\Sigma_b^-)
	+\mathcal{M}(\rho^+,\Lambda_b^0;\Sigma_b^-)\non
	&+\mathcal{M}(\rho^+,\Sigma_b^0;\Sigma_b^-)+\mathcal{M}(K^+,\Xi_b^0;\Xi_b^-)+\mathcal{M}(K^+,\Xi_b^0;\Xi_b^{\prime -})+\mathcal{M}(K^+,\Xi_b^{\prime 0};\Xi_b^-)
	+\mathcal{M}(K^+,\Xi_b^{\prime 0};\Xi_b^{\prime -})\non
	&+\mathcal{M}(K^{\ast +},\Xi_b^0;\Xi_b^-)+\mathcal{M}(K^{\ast +},\Xi_b^0;\Xi_b^{\prime -})+\mathcal{M}(K^{\ast +},\Xi_b^{\prime 0};\Xi_b^-)
	+\mathcal{M}(K^{\ast +},\Xi_b^{\prime 0};\Xi_b^{\prime -}), \\
	\mathcal{A}(\Xi_{bc}^+\to\Sigma_b^+\eta_8)=&\mathcal{C}_{SD}(\Xi_{bc}^+\to\Sigma_b^+\eta_8)+\mathcal{M}(\pi^+,\Sigma_b^0;\Sigma_b^0)+\mathcal{M}(\pi^+,\Lambda_b^0;\Lambda_b^0)+\mathcal{M}(\rho^+,\Sigma_b^0;\Sigma_b^0)
	+\mathcal{M}(\rho^+,\Lambda_b^0;\Lambda_b^0)\non
	&+\mathcal{M}(K^+,\Xi_b^0;K^{\ast +})+\mathcal{M}(K^+,\Xi_b^{\prime 0};K^{\ast +})+\mathcal{M}(K^{\ast +},\Xi_b^0;K^{+})+\mathcal{M}(K^{\ast +},\Xi_b^{\prime 0};K^{+})
	+\mathcal{M}(K^+,\Xi_b^0;\Xi_b^0)\non
	&+\mathcal{M}(K^+,\Xi_b^0;\Xi_b^{\prime 0})+\mathcal{M}(K^+,\Xi_b^{\prime 0};\Xi_b^0)+\mathcal{M}(K^+,\Xi_b^{\prime 0};\Xi_b^{\prime 0})+\mathcal{M}(K^{\ast +},\Xi_b^0;\Xi_b^0)
	+\mathcal{M}(K^{\ast +},\Xi_b^0;\Xi_b^{\prime 0})\non
	&+\mathcal{M}(K^{\ast +},\Xi_b^{\prime 0};\Xi_b^0)+\mathcal{M}(K^{\ast +},\Xi_b^{\prime 0};\Xi_b^{\prime 0}),\\
	\mathcal{A}(\Xi_{bc}^+\to\Sigma_b^+\bar{K}^0)=&\mathcal{C}_{SD}(\Xi_{bc}^+\to\Sigma_b^+\bar{K}^0)+\mathcal{M}(\pi^+,\Xi_b^0;K^{\ast +})+\mathcal{M}(\pi^+,\Xi_b^{\prime 0};K^{\ast +})+\mathcal{M}(\rho^+,\Xi_b^0;K^{+})
	+\mathcal{M}(\rho^+,\Xi_b^{\prime 0};K^{+})\non
	&+\mathcal{M}(\pi^+,\Xi_b^0;\Sigma_b^0)+\mathcal{M}(\pi^+,\Xi_b^0;\Lambda_b^0)+\mathcal{M}(\pi^+,\Xi_b^{\prime 0};\Sigma_b^0)+\mathcal{M}(\pi^+,\Xi_b^{\prime 0};\Lambda_b^0)
	+\mathcal{M}(\rho^+,\Xi_b^0;\Sigma_b^0)\non
	&+\mathcal{M}(\rho^+,\Xi_b^0;\Lambda_b^0)+\mathcal{M}(\rho^+,\Xi_b^{\prime 0};\Sigma_b^0)+\mathcal{M}(\rho^+,\Xi_b^{\prime 0};\Lambda_b^0),\\
	\mathcal{A}(\Xi_{bc}^+\to\Sigma_b^+K^0)=&\mathcal{C}_{SD}(\Xi_{bc}^+\to\Sigma_b^+K^0)+\mathcal{M}(K^{\ast +},\Lambda_b^0;\pi^+)+\mathcal{M}(K^{\ast +},\Sigma_b^0;\pi^+)+\mathcal{M}(K^{+},\Lambda_b^0;\rho^+)
	+\mathcal{M}(K^{+},\Sigma_b^0;\rho^+)\non
	&+\mathcal{M}(K^{+},\Lambda_b^0;\Xi_b^0)+\mathcal{M}(K^{+},\Lambda_b^0;\Xi_b^{\prime 0})+\mathcal{M}(K^{+},\Sigma_b^0;\Xi_b^0)+\mathcal{M}(K^{+},\Sigma_b^0;\Xi_b^{\prime 0})
	+\mathcal{M}(K^{\ast +},\Lambda_b^0;\Xi_b^0)\non
	&+\mathcal{M}(K^{\ast +},\Lambda_b^0;\Xi_b^{\prime 0})+\mathcal{M}(K^{\ast +},\Sigma_b^0;\Xi_b^0)+\mathcal{M}(K^{\ast +},\Sigma_b^0;\Xi_b^{\prime 0}), \\
	\mathcal{A}(\Xi_{bc}^+\to\Sigma_b^+\pi^0)=&\mathcal{C}_{SD}(\Xi_{bc}^+\to\Sigma_b^+\pi^0)+\mathcal{M}(\pi^+,\Lambda_b^0;\rho^+)+\mathcal{M}(\pi^+,\Sigma_b^0;\rho^+)+\mathcal{M}(\rho^+,\Lambda_b^0;\pi^+)
	+\mathcal{M}(\rho^+,\Sigma_b^0;\pi^+)\non
	&+\mathcal{M}(\pi^+,\Lambda_b^0;\Sigma^+)+\mathcal{M}(\pi^+,\Sigma_b^0;\Lambda^+)+\mathcal{M}(\rho^+,\Lambda_b^0;\Sigma^+)+\mathcal{M}(\rho^+,\Sigma_b^0;\Lambda^+)
	+\mathcal{M}(K^+,\Xi_b^0;K^{\ast +})\non
	&+\mathcal{M}(K^+,\Xi_b^{\prime 0};K^{\ast +})+\mathcal{M}(K^{\ast +},\Xi_b^0;K^{+})+\mathcal{M}(K^{\ast +},\Xi_b^{\prime 0};K^{+})+\mathcal{M}(K^{+},\Xi_b^0;\Xi_b^0)
	+\mathcal{M}(K^{+},\Xi_b^0;\Xi_b^{\prime 0})\non
	&+\mathcal{M}(K^{+},\Xi_b^{\prime 0};\Xi_b^0)+\mathcal{M}(K^{+},\Xi_b^{\prime 0};\Xi_b^{\prime 0})+\mathcal{M}(K^{\ast +},\Xi_b^0;\Xi_b^0)+\mathcal{M}(K^{\ast +},\Xi_b^0;\Xi_b^{\prime 0})
	+\mathcal{M}(K^{\ast +},\Xi_b^{\prime 0};\Xi_b^0)\non
	&+\mathcal{M}(K^{\ast +},\Xi_b^{\prime 0};\Xi_b^{\prime 0}),\\
	\mathcal{A}(\Xi_{bc}^+\to\Xi_b^{\prime 0}K^+)=&\mathcal{T}(\Xi_{bc}^+\to\Xi_b^{\prime 0}K^+)+\mathcal{M}(K^{+},\Xi_b^0;\rho^0)+\mathcal{M}(K^{+},\Xi_b^0;\omega)+\mathcal{M}(K^{+},\Xi_b^{\prime 0};\rho^0)
	+\mathcal{M}(K^{+},\Xi_b^{\prime 0};\omega)\non
	&+\mathcal{M}(K^{\ast +},\Xi_b^0;\pi^0)+\mathcal{M}(K^{\ast +},\Xi_b^0;\eta_8)+\mathcal{M}(K^{\ast +},\Xi_b^{\prime 0};\pi^0)+\mathcal{M}(K^{\ast +},\Xi_b^{\prime 0};\eta_8)
	+\mathcal{M}(\pi^+,\Lambda_b^0;\bar{K}^{\ast 0})\non
	&+\mathcal{M}(\pi^+,\Sigma_b^0;\bar{K}^{\ast 0})+\mathcal{M}(\rho^+,\Lambda_b^0;\bar{K}^{0})+\mathcal{M}(\rho^+,\Sigma_b^0;\bar{K}^{0})+\mathcal{M}(K^+,\Xi_b^0;\phi)
	+\mathcal{M}(K^+,\Xi_b^{\prime 0};\phi)\non
	&+\mathcal{M}(K^+,\Xi_b^0;\Omega_b^-)+\mathcal{M}(K^+,\Xi_b^{\prime 0};\Omega_b^-)+\mathcal{M}(K^{\ast +},\Xi_b^0;\Omega_b^-)+\mathcal{M}(K^{\ast +},\Xi_b^{\prime 0};\Omega_b^-)
	+\mathcal{M}(\pi^+,\Lambda_b^0;\Xi_b^-)\non
	&+\mathcal{M}(\pi^+,\Lambda_b^0;\Xi_b^{\prime -})+\mathcal{M}(\pi^+,\Sigma_b^0;\Xi_b^-)+\mathcal{M}(\pi^+,\Sigma_b^0;\Xi_b^{\prime -})+\mathcal{M}(\rho^+,\Lambda_b^0;\Xi_b^-)
	+\mathcal{M}(\rho^+,\Lambda_b^0;\Xi_b^{\prime -})\non
	&+\mathcal{M}(\rho^+,\Sigma_b^0;\Xi_b^-)+\mathcal{M}(\rho^+,\Sigma_b^0;\Xi_b^{\prime -}),\\
	\mathcal{A}(\Xi_{bc}^+\to\Xi_b^{\prime 0}\pi^+)=&\mathcal{T}(\Xi_{bc}^+\to\Xi_b^{\prime 0}\pi^+)+\mathcal{M}(\pi^+,\Xi_b^0;\rho^0)+\mathcal{M}(\pi^+,\Xi_b^{\prime 0};\rho^0)+\mathcal{M}(\rho^+,\Xi_b^0;\pi^0)
	+\mathcal{M}(\rho^+,\Xi_b^{\prime 0};\pi^0)\non
	&+\mathcal{M}(\pi^+,\Xi_b^0;\Xi_b^-)+\mathcal{M}(\pi^+,\Xi_b^0;\Xi_b^{\prime -})+\mathcal{M}(\pi^+,\Xi_b^{\prime 0};\Xi_b^-)+\mathcal{M}(\pi^+,\Xi_b^{\prime 0};\Xi_b^{\prime -})
	+\mathcal{M}(\rho^+,\Xi_b^0;\Xi_b^-)\non
	&+\mathcal{M}(\rho^+,\Xi_b^0;\Xi_b^{\prime -})+\mathcal{M}(\rho^+,\Xi_b^{\prime 0};\Xi_b^-)+\mathcal{M}(\rho^+,\Xi_b^{\prime 0};\Xi_b^{\prime -}) ,\\
	\mathcal{A}(\Xi_{bc}^0\to\Omega_b^-K^+)=&\mathcal{M}(\pi^+,\Xi_b^-;\bar{K}^{\ast 0})+\mathcal{M}(\pi^+,\Xi_b^{\prime -};\bar{K}^{\ast 0})+\mathcal{M}(\rho^+,\Xi_b^-;\bar{K}^{0})
	+\mathcal{M}(\rho^+,\Xi_b^{\prime -};\bar{K}^{0}), \\
	\mathcal{A}(\Xi_{bc}^0\to\Sigma_b^0\eta_8)=&\mathcal{C}_{SD}(\Xi_{bc}^0\to\Sigma_b^0\eta_8)+\mathcal{M}(K^+,\Xi_b^-;K^{\ast +})+\mathcal{M}(K^+,\Xi_b^{\prime -};K^{\ast +})+\mathcal{M}(K^{\ast +},\Xi_b^-;K^{+})
	+\mathcal{M}(K^{\ast +},\Xi_b^{\prime -};K^{+})\non
	&+\mathcal{M}(K^{+},\Xi_b^-;\Xi_b^-)+\mathcal{M}(K^{+},\Xi_b^-;\Xi_b^{\prime -})+\mathcal{M}(K^{+},\Xi_b^{\prime -};\Xi_b^-)+\mathcal{M}(K^{+},\Xi_b^{\prime -};\Xi_b^{\prime -})
	+\mathcal{M}(K^{\ast +},\Xi_b^-;\Xi_b^-)\non
	&+\mathcal{M}(K^{\ast +},\Xi_b^-;\Xi_b^{\prime -})+\mathcal{M}(K^{\ast +},\Xi_b^{\prime -};\Xi_b^-)+\mathcal{M}(K^{\ast +},\Xi_b^{\prime -};\Xi_b^{\prime -})+\mathcal{M}(\pi^+,\Sigma_b^-;\Sigma_b^-)
	+\mathcal{M}(\rho^+,\Sigma_b^-;\Sigma_b^-), \\
	\mathcal{A}(\Xi_{bc}^0\to\Sigma_b^0\bar{K}^0)=&\mathcal{C}_{SD}(\Xi_{bc}^0\to\Sigma_b^0\bar{K}^0)+\mathcal{M}(\pi^+,\Xi_b^-;K^{\ast +})+\mathcal{M}(\pi^+,\Xi_b^{\prime -};K^{\ast +})+\mathcal{M}(\rho^+,\Xi_b^-;K^{+})
	+\mathcal{M}(\rho^+,\Xi_b^{\prime -};K^{+})\non
	&+\mathcal{M}(\pi^+,\Xi_b^-;\Sigma_b^-)+\mathcal{M}(\pi^+,\Xi_b^{\prime -};\Sigma_b^-)+\mathcal{M}(\rho^+,\Xi_b^-;\Sigma_b^-)
	+\mathcal{M}(\rho^+,\Xi_b^{\prime -};\Sigma_b^-),\\
	\mathcal{A}(\Xi_{bc}^0\to\Sigma_b^0K^0)=&\mathcal{C}_{SD}(\Xi_{bc}^0\to\Sigma_b^0K^0)+\mathcal{M}(K^+,\Sigma_b^-;\rho^+)+\mathcal{M}(K^{\ast +},\Sigma_b^-;\pi^+)+\mathcal{M}(K^+,\Sigma_b^-;\Xi_b^-)
	+\mathcal{M}(K^+,\Sigma_b^-;\Xi_b^{\prime -})\non
	&+\mathcal{M}(K^{\ast +},\Sigma_b^-;\Xi_b^-)+\mathcal{M}(K^{\ast +},\Sigma_b^-;\Xi_b^{\prime -}),\\
	\mathcal{A}(\Xi_{bc}^0\to\Sigma_b^0\pi^0)=&\mathcal{C}_{SD}(\Xi_{bc}^0\to\Sigma_b^0\pi^0)+\mathcal{M}(\pi^+,\Sigma_b^-;\rho^+)+\mathcal{M}(\rho^+,\Sigma_b^-;\pi^+)+\mathcal{M}(K^+,\Xi_b^-;K^{\ast +})
	+\mathcal{M}(K^+,\Xi_b^{\prime -};K^{\ast +})\non
	&+\mathcal{M}(K^{\ast +},\Xi_b^-;K^{+})+\mathcal{M}(K^{\ast +},\Xi_b^{\prime -};K^{+})+\mathcal{M}(\pi^+,\Sigma_b^-;\Sigma_b^-)+\mathcal{M}(\rho^+,\Sigma_b^-;\Sigma_b^-)
	+\mathcal{M}(K^+,\Xi_b^-;\Xi_b^-)\non
	&+\mathcal{M}(K^+,\Xi_b^-;\Xi_b^{\prime -})+\mathcal{M}(K^+,\Xi_b^{\prime -};\Xi_b^-)+\mathcal{M}(K^+,\Xi_b^{\prime -};\Xi_b^{\prime -})+\mathcal{M}(K^{\ast +},\Xi_b^-;\Xi_b^-)
	+\mathcal{M}(K^{\ast +},\Xi_b^-;\Xi_b^{\prime -})\non
	&+\mathcal{M}(K^{\ast +},\Xi_b^{\prime -};\Xi_b^-)+\mathcal{M}(K^{\ast +},\Xi_b^{\prime -};\Xi_b^{\prime -}),\\
	\mathcal{A}(\Xi_{bc}^0\to\Sigma_b^-\pi^+)=&\mathcal{T}(\Xi_{bc}^0\to\Sigma_b^-\pi^+)+\mathcal{M}(\pi^+,\Sigma_b^-;\rho^0)+\mathcal{M}(\rho^+,\Sigma_b^-;\pi^0)+\mathcal{M}(K^+,\Xi_b^-;K^{\ast 0})
	+\mathcal{M}(K^+,\Xi_b^{\prime -};K^{\ast 0})\non
	&+\mathcal{M}(K^{\ast +},\Xi_b^-;K^{0})+\mathcal{M}(K^{\ast +},\Xi_b^{\prime -};K^{0}),\\
	\mathcal{A}(\Xi_{bc}^0\to\Sigma_b^+K^-)=&\mathcal{M}(\pi^+,\Xi_b^-;\Sigma_b^0)+\mathcal{M}(\pi^+,\Xi_b^-;\Lambda_b^0)+\mathcal{M}(\pi^+,\Xi_b^{\prime -};\Sigma_b^0)+\mathcal{M}(\pi^+,\Xi_b^{\prime -};\Lambda_b^0)
	+\mathcal{M}(\rho^+,\Xi_b^-;\Sigma_b^0)\non
	&+\mathcal{M}(\rho^+,\Xi_b^-;\Lambda_b^0)+\mathcal{M}(\rho^+,\Xi_b^{\prime -};\Sigma_b^0)
	+\mathcal{M}(\rho^+,\Xi_b^{\prime -};\Lambda_b^0),\\
	\mathcal{A}(\Xi_{bc}^0\to\Sigma_b^+\pi^-)=&\mathcal{M}(\pi^+,\Sigma_b^{-};\Sigma_b^0)+\mathcal{M}(\pi^+,\Sigma_b^{-};\Lambda_b^0)+\mathcal{M}(\rho^+,\Sigma_b^{-};\Sigma_b^0)+\mathcal{M}(\rho^+,\Sigma_b^{-};\Lambda_b^0)
	+\mathcal{M}(K^+,\Xi_b^{-};\Xi_b^0)\non
	&+\mathcal{M}(K^+,\Xi_b^{-};\Xi_b^{\prime 0})+\mathcal{M}(K^+,\Xi_b^{\prime -};\Xi_b^0)+\mathcal{M}(K^+,\Xi_b^{\prime -};\Xi_b^{\prime 0})
	+\mathcal{M}(K^{\ast +},\Xi_b^{-};\Xi_b^0)\non
	&+\mathcal{M}(K^{\ast +},\Xi_b^{-};\Xi_b^{\prime 0})+\mathcal{M}(K^{\ast +},\Xi_b^{\prime -};\Xi_b^0)
	+\mathcal{M}(K^{\ast +},\Xi_b^{\prime -};\Xi_b^{\prime 0}), \\
	\mathcal{A}(\Xi_{bc}^0\to\Xi_b^{\prime -}K^+)=&\mathcal{T}(\Xi_{bc}^0\to\Xi_b^{\prime -}K^+)+\mathcal{M}(\pi^+,\Sigma_b^{-};\bar{K}^{\ast 0})+\mathcal{M}(\rho^+,\Sigma_b^{-};\bar{K}^{0})+\mathcal{M}(K^+,\Xi_b^{-};\phi)
	+\mathcal{M}(K^+,\Xi_b^{\prime -};\phi)\non
	&+\mathcal{M}(K^{\ast +},\Xi_b^{-};\eta_8)+\mathcal{M}(K^{\ast +},\Xi_b^{\prime -};\eta_8),\\
	\mathcal{A}(\Xi_{bc}^0\to\Xi_b^{\prime -}\pi^+)=&\mathcal{T}(\Xi_{bc}^0\to\Xi_b^{\prime -}\pi^+)+\mathcal{M}(\pi^+,\Xi_b^{-};\rho^0)+\mathcal{M}(\pi^+,\Xi_b^{\prime -};\rho^0)+\mathcal{M}(\rho^+,\Xi_b^{-};\pi^0)
	+\mathcal{M}(\rho^+,\Xi_b^{\prime -};\pi^0),\\
	\mathcal{A}(\Xi_{bc}^0\to\Xi_b^{\prime 0}\eta_8)=&\mathcal{C}_{SD}(\Xi_{bc}^0\to\Xi_b^{\prime 0}\eta_8)+\mathcal{M}(\pi^+,\Xi_b^{-};\Xi_b^-)+\mathcal{M}(\pi^+,\Xi_b^{-};\Xi_b^{\prime -})+\mathcal{M}(\pi^+,\Xi_b^{\prime -};\Xi_b^-)
	+\mathcal{M}(\pi^+,\Xi_b^{\prime -};\Xi_b^{\prime -})\non
	&+\mathcal{M}(\rho^+,\Xi_b^{-};\Xi_b^-)+\mathcal{M}(\rho^+,\Xi_b^{-};\Xi_b^{\prime -})+\mathcal{M}(\rho^+,\Xi_b^{\prime -};\Xi_b^-)
	+\mathcal{M}(\rho^+,\Xi_b^{\prime -};\Xi_b^{\prime -}),\\
	\mathcal{A}(\Xi_{bc}^0\to\Xi_b^{\prime 0}K^0)=&\mathcal{C}_{SD}(\Xi_{bc}^0\to\Xi_b^{\prime 0}K^0)+\mathcal{M}(K^+,\Xi_b^{-};\rho^+)+\mathcal{M}(K^+,\Xi_b^{\prime -};\rho^+)+\mathcal{M}(K^{\ast +},\Xi_b^{-};\pi^+)
	+\mathcal{M}(K^{\ast +},\Xi_b^{\prime -};\pi^+)\non
	&+\mathcal{M}(K^+,\Xi_b^{-};\Omega_b^-)+\mathcal{M}(K^+,\Xi_b^{\prime -};\Omega_b^-)+\mathcal{M}(K^{\ast +},\Xi_b^{-};\Omega_b^-)+\mathcal{M}(K^{\ast +},\Xi_b^{\prime -};\Omega_b^-)
	+\mathcal{M}(\pi^+,\Sigma_b^{-};\Xi_b^-)\non
	&+\mathcal{M}(\pi^+,\Sigma_b^{-};\Xi_b^{\prime -})+\mathcal{M}(\rho^+,\Sigma_b^{-};\Xi_b^-)
	+\mathcal{M}(\rho^+,\Sigma_b^{-};\Xi_b^{\prime -}), \\
	\mathcal{A}(\Xi_{bc}^0\to\Xi_b^{\prime 0}\pi^0)=&\mathcal{C}_{SD}(\Xi_{bc}^0\to\Xi_b^{\prime 0}\pi^0)+\mathcal{M}(\pi^+,\Xi_b^{-};\rho^+)+\mathcal{M}(\pi^+,\Xi_b^{\prime -};\rho^+)+\mathcal{M}(\rho^+,\Xi_b^{-};\pi^+)
	+\mathcal{M}(\rho^+,\Xi_b^{\prime -};\pi^+)\non
	&+\mathcal{M}(\pi^+,\Xi_b^{-};\Xi_b^-)+\mathcal{M}(\pi^+,\Xi_b^{-};\Xi_b^{\prime -})+\mathcal{M}(\pi^+,\Xi_b^{\prime -};\Xi_b^-)+\mathcal{M}(\pi^+,\Xi_b^{\prime -};\Xi_b^{\prime -})
	+\mathcal{M}(\rho^+,\Xi_b^{-};\Xi_b^-)\non
	&+\mathcal{M}(\rho^+,\Xi_b^{-};\Xi_b^{\prime -})+\mathcal{M}(\rho^+,\Xi_b^{\prime -};\Xi_b^-)
	+\mathcal{M}(\rho^+,\Xi_b^{\prime -};\Xi_b^{\prime -}),\\
	\mathcal{A}(\Omega_{bc}^0\to\Omega_b^-K^+)=&\mathcal{T}(\Omega_{bc}^0\to\Omega_b^-K^+)+\mathcal{M}(\pi^+,\Xi_b^{-};\bar{K}^{\ast 0})+\mathcal{M}(\pi^+,\Xi_b^{\prime -};\bar{K}^{\ast 0})+\mathcal{M}(\rho^+,\Xi_b^{-};\bar{K}^{0})
	+\mathcal{M}(\rho^+,\Xi_b^{\prime -};\bar{K}^{0})\non
	&+\mathcal{M}(K^+,\Omega_b^{-};\phi)+\mathcal{M}(K^{\ast +},\Omega_b^{-};\eta_8),\\
	\mathcal{A}(\Omega_{bc}^0\to\Sigma_b^0\eta_8)=&\mathcal{C}_{SD}(\Omega_{bc}^0\to\Sigma_b^0\eta_8)+\mathcal{M}(K^+,\Xi_b^{-};K^{\ast +})+\mathcal{M}(K^+,\Xi_b^{\prime -};K^{\ast +})+\mathcal{M}(K^{\ast +},\Xi_b^{-};K^{+})
	+\mathcal{M}(K^{\ast +},\Xi_b^{\prime -};K^{+})\non
	&+\mathcal{M}(K^{+},\Xi_b^{-};\Xi_b^-)+\mathcal{M}(K^{+},\Xi_b^{-};\Xi_b^{\prime -})+\mathcal{M}(K^{+},\Xi_b^{\prime -};\Xi_b^-)+\mathcal{M}(K^{+},\Xi_b^{\prime -};\Xi_b^{\prime -})
	+\mathcal{M}(K^{\ast +},\Xi_b^{-};\Xi_b^-)\non
	&+\mathcal{M}(K^{\ast +},\Xi_b^{-};\Xi_b^{\prime -})+\mathcal{M}(K^{\ast +},\Xi_b^{\prime -};\Xi_b^-)
	+\mathcal{M}(K^{\ast +},\Xi_b^{\prime -};\Xi_b^{\prime -}),\\
	\mathcal{A}(\Omega_{bc}^0\to\Sigma_b^0\bar{K}^0)=&\mathcal{C}_{SD}(\Omega_{bc}^0\to\Sigma_b^0\bar{K}^0)+\mathcal{M}(\pi^+,\Xi_b^{-};K^{\ast +})+\mathcal{M}(\pi^+,\Xi_b^{\prime -};K^{\ast +})+\mathcal{M}(\rho^+,\Xi_b^{-};K^{+})
	+\mathcal{M}(\rho^+,\Xi_b^{\prime -};K^{+})\non
	&+\mathcal{M}(\pi^+,\Xi_b^{-};\Sigma_b^-)+\mathcal{M}(\pi^+,\Xi_b^{\prime -};\Sigma_b^-)+\mathcal{M}(\rho^+,\Xi_b^{-};\Sigma_b^-)+\mathcal{M}(\rho^+,\Xi_b^{\prime -};\Sigma_b^-)
	+\mathcal{M}(K^+,\Omega_b^{-};\Xi_b^-)\non
	&+\mathcal{M}(K^+,\Omega_b^{-};\Xi_b^{\prime -})+\mathcal{M}(K^{\ast +},\Omega_b^{-};\Xi_b^-)+\mathcal{M}(K^{\ast +},\Omega_b^{-};\Xi_b^{\prime -}),\\
	\mathcal{A}(\Omega_{bc}^0\to\Sigma_b^0\pi^0)=&\mathcal{M}(K^{+},\Xi_b^{-};K^{\ast +})+\mathcal{M}(K^{+},\Xi_b^{\prime -};K^{\ast +})+\mathcal{M}(K^{\ast +},\Xi_b^{-};K^{+})+\mathcal{M}(K^{\ast +},\Xi_b^{\prime -};K^{+})
	+\mathcal{M}(K^{+},\Xi_b^{-};\Xi_b^-)\non
	&+\mathcal{M}(K^{+},\Xi_b^{-};\Xi_b^{\prime -})+\mathcal{M}(K^{+},\Xi_b^{\prime -};\Xi_b^-)+\mathcal{M}(K^{+},\Xi_b^{\prime -};\Xi_b^{\prime -})+\mathcal{M}(K^{\ast +},\Xi_b^{-};\Xi_b^-)
	+\mathcal{M}(K^{\ast +},\Xi_b^{-};\Xi_b^{\prime -})\non
	&+\mathcal{M}(K^{\ast +},\Xi_b^{\prime -};\Xi_b^-)+\mathcal{M}(K^{\ast +},\Xi_b^{\prime -};\Xi_b^{\prime -}), \\
	\mathcal{A}(\Omega_{bc}^0\to\Sigma_b^-\pi^+)=&\mathcal{M}(K^{+},\Xi_b^{-};K^{\ast 0})+\mathcal{M}(K^{+},\Xi_b^{\prime -};K^{\ast 0})+\mathcal{M}(K^{\ast +},\Xi_b^{-};K^{0})
	+\mathcal{M}(K^{\ast +},\Xi_b^{\prime -};K^{0}),\\
	\mathcal{A}(\Omega_{bc}^0\to\Sigma_b^+K^-)=&\mathcal{M}(\pi^+,\Xi_b^{-};\Sigma_b^0)+\mathcal{M}(\pi^+,\Xi_b^{-};\Lambda_b^0)+\mathcal{M}(\pi^+,\Xi_b^{\prime -};\Sigma_b^0)+\mathcal{M}(\pi^+,\Xi_b^{\prime -};\Lambda_b^0)
	+\mathcal{M}(\rho^+,\Xi_b^{-};\Sigma_b^0)\non
	&+\mathcal{M}(\rho^+,\Xi_b^{-};\Lambda_b^0)+\mathcal{M}(\rho^+,\Xi_b^{\prime -};\Sigma_b^0)+\mathcal{M}(\rho^+,\Xi_b^{\prime -};\Lambda_b^0)+\mathcal{M}(K^+,\Omega_b^{-};\Xi_b^0)
	+\mathcal{M}(K^+,\Omega_b^{-};\Xi_b^{\prime 0})\non
	&+\mathcal{M}(K^{\ast +},\Omega_b^{-};\Xi_b^0)+\mathcal{M}(K^{\ast +},\Omega_b^{-};\Xi_b^{\prime 0}),\\
	\mathcal{A}(\Omega_{bc}^0\to\Sigma_b^+\pi^-)=&\mathcal{M}(K^{+},\Xi_b^{-};\Xi_b^0)+\mathcal{M}(K^{+},\Xi_b^{-};\Xi_b^{\prime 0})+\mathcal{M}(K^{+},\Xi_b^{\prime -};\Xi_b^0)+\mathcal{M}(K^{+},\Xi_b^{\prime -};\Xi_b^{\prime 0})
	+\mathcal{M}(K^{\ast +},\Xi_b^{-};\Xi_b^0)\non
	&+\mathcal{M}(K^{\ast +},\Xi_b^{-};\Xi_b^{\prime 0})+\mathcal{M}(K^{\ast +},\Xi_b^{\prime -};\Xi_b^0)
	+\mathcal{M}(K^{\ast +},\Xi_b^{\prime -};\Xi_b^{\prime 0}),\\
	\mathcal{A}(\Omega_{bc}^0\to\Xi_b^{\prime -}K^+)=&\mathcal{T}(\Omega_{bc}^0\to\Xi_b^{\prime -}K^+)+\mathcal{M}(K^{+},\Xi_b^{-};\phi)+\mathcal{M}(K^{+},\Xi_b^{\prime -};\phi)+\mathcal{M}(K^{\ast +},\Xi_b^{-};\eta_8)
	+\mathcal{M}(K^{\ast +},\Xi_b^{\prime -};\eta_8), \\
	\mathcal{A}(\Omega_{bc}^0\to\Xi_b^{\prime -}\pi^+)=&\mathcal{T}(\Omega_{bc}^0\to\Xi_b^{\prime -}\pi^+)+\mathcal{M}(\pi^+,\Xi_b^{-};\rho^0)+\mathcal{M}(\pi^+,\Xi_b^{\prime -};\rho^0)+\mathcal{M}(\rho^+,\Xi_b^{-};\pi^0)
	+\mathcal{M}(\rho^+,\Xi_b^{\prime -};\pi^0)\non
	&+\mathcal{M}(K^+,\Omega_b^{-};K^{\ast 0})+\mathcal{M}(K^{\ast +},\Omega_b^{-};K^{0}),\\
	\mathcal{A}(\Omega_{bc}^0\to\Xi_b^{\prime 0}\eta_8)=&\mathcal{C}_{SD}(\Omega_{bc}^0\to\Xi_b^{\prime 0}\eta_8)+\mathcal{M}(K^+,\Omega_b^{-};K^{\ast +})+\mathcal{M}(K^{\ast +},\Omega_b^{-};K^{+})+\mathcal{M}(K^+,\Omega_b^{-};\Omega_b^-)
	+\mathcal{M}(K^{\ast +},\Omega_b^{-};\Omega_b^-)\non
	&+\mathcal{M}(\pi^+,\Xi_b^{-};\Xi_b^-)+\mathcal{M}(\pi^+,\Xi_b^{-};\Xi_b^{\prime -})+\mathcal{M}(\pi^+,\Xi_b^{\prime -};\Xi_b^-)+\mathcal{M}(\pi^+,\Xi_b^{\prime -};\Xi_b^{\prime -})
	+\mathcal{M}(\rho^+,\Xi_b^{-};\Xi_b^-)\non
	&+\mathcal{M}(\rho^+,\Xi_b^{-};\Xi_b^{\prime -})+\mathcal{M}(\rho^+,\Xi_b^{\prime -};\Xi_b^-)+\mathcal{M}(\rho^+,\Xi_b^{\prime -};\Xi_b^{\prime -})+\mathcal{M}(\pi^+,\Xi_b^{-};\rho^+)
	+\mathcal{M}(\pi^+,\Xi_b^{\prime -};\rho^+)\non
	&+\mathcal{M}(\rho^+,\Xi_b^{-};\pi^+)+\mathcal{M}(\rho^+,\Xi_b^{\prime -};\pi^+),\\
	\mathcal{A}(\Omega_{bc}^0\to\Xi_b^{\prime 0}\bar{K}^0)=&\mathcal{C}_{SD}(\Omega_{bc}^0\to\Xi_b^{\prime 0}\bar{K}^0)+\mathcal{M}(\pi^+,\Omega_b^{-};K^{\ast +})+\mathcal{M}(\rho^+,\Omega_b^{-};K^{+})+\mathcal{M}(\pi^+,\Omega_b^{-};\Xi_b^-)
	+\mathcal{M}(\pi^+,\Omega_b^{-};\Xi_b^{\prime -})\non
	&+\mathcal{M}(\rho^+,\Omega_b^{-};\Xi_b^-)\mathcal{M}(\rho^+,\Omega_b^{-};\Xi_b^{\prime -}),\\
	\mathcal{A}(\Omega_{bc}^0\to\Xi_b^{\prime 0}K^0)=&\mathcal{C}_{SD}(\Omega_{bc}^0\to\Xi_b^{\prime 0}K^0)+\mathcal{M}(K^+,\Xi_b^{-};\rho^+)+\mathcal{M}(K^+,\Xi_b^{\prime -};\rho^+)+\mathcal{M}(K^{\ast +},\Xi_b^{-};\pi^+)
	+\mathcal{M}(K^{\ast +},\Xi_b^{\prime -};\pi^+)\non
	&+\mathcal{M}(K^{+},\Xi_b^{-};\Omega_b^-)+\mathcal{M}(K^{+},\Xi_b^{\prime -};\Omega_b^-)+\mathcal{M}(K^{\ast +},\Xi_b^{-};\Omega_b^-)
	+\mathcal{M}(K^{\ast +},\Xi_b^{\prime -};\Omega_b^-),\\
	\mathcal{A}(\Omega_{bc}^0\to\Xi_b^{\prime 0}\pi^0)=&\mathcal{C}_{SD}(\Omega_{bc}^0\to\Xi_b^{\prime 0}\pi^0)+\mathcal{M}(\pi^+,\Xi_b^{-};\rho^+)+\mathcal{M}(\pi^+,\Xi_b^{\prime -};\rho^+)+\mathcal{M}(\rho^+,\Xi_b^{-};\pi^+)
	+\mathcal{M}(\rho^+,\Xi_b^{\prime -};\pi^+)\non
	&+\mathcal{M}(K^+,\Omega_b^{-};K^{\ast +})+\mathcal{M}(K^{\ast +},\Omega_b^{-};K^{+})+\mathcal{M}(\pi^+,\Xi_b^{-};\Xi_b^-)+\mathcal{M}(\pi^+,\Xi_b^{-};\Xi_b^{\prime -})
	+\mathcal{M}(\pi^+,\Xi_b^{\prime -};\Xi_b^-)\non
	&+\mathcal{M}(\pi^+,\Xi_b^{\prime -};\Xi_b^{\prime -})+\mathcal{M}(\rho^+,\Xi_b^{-};\Xi_b^-)+\mathcal{M}(\rho^+,\Xi_b^{-};\Xi_b^{\prime -})+\mathcal{M}(\rho^+,\Xi_b^{\prime -};\Xi_b^-)
	+\mathcal{M}(\rho^+,\Xi_b^{\prime -};\Xi_b^{\prime -}).
	\end{align}}

\end{document}